\documentclass[article]{jss}

\usepackage{orcidlink,thumbpdf,lmodern}

\usepackage{framed}
\usepackage{amsmath,amssymb,array}
\usepackage{booktabs}
\usepackage{float}
\usepackage{pifont}

\newcommand{\cmark}{\ding{51}}%

\author{Nicoletta D'Angelo~\orcidlink{0000-0002-8878-5986}\\University of Palermo
   \And Giada Adelfio~\orcidlink{0000-0002-3194-4296}\\University of Palermo}
\Plainauthor{Nicoletta D'Angelo, Giada Adelfio}

\title{\pkg{stopp}: An \proglang{R} Package for Spatio-Temporal Point Pattern Analysis}
\Plaintitle{stopp: An R Package for Spatio-Temporal Point Pattern Analysis}

\Abstract{
\pkg{stopp} is a novel \proglang{R} package specifically designed for the analysis of spatio-temporal point patterns which might have occurred in a subset of the Euclidean space or on some specific linear network, such as roads of a city.
It represents the first package providing a comprehensive modelling framework for spatio-temporal Poisson point processes. 
While many specialized models exist in the scientific literature for analyzing complex spatio-temporal point patterns, we address the lack of general software for comparing simpler alternative models and their goodness of fit.
The package's main functionalities include modelling and diagnostics, together with exploratory analysis tools and the simulation of point processes. A particular focus is given to local first-order and second-order characteristics. The package aggregates existing methods
within one coherent framework, including those we proposed in recent papers, and it aims to welcome many further proposals and extensions from the \proglang{R} community.
}

\Plainkeywords{point patterns, simulation, model fitting, diagnostics, local analyses,  spatial statistics, space-time point
	processes, R}

\Address{
  Nicoletta D'Angelo and Giada Adelfio\\
  Department of Economics, Business, and Statistics\\
  University of Palermo\\
  Palermo, Italy\\
  E-mail: \email{nicoletta.dangelo@unipa.it}
}

\begin{document}



\section{Introduction}

\pkg{stopp} \citep{stopp} is a new package in the \proglang{R} language for analysing point patterns in three dimensions. 
The first two dimensions represent spatial components, while the third dimension is regarded as temporal.
The \pkg{stopp} package has been published on the Comprehensive R Archive Network (CRAN) and is available from \url{https://CRAN.R-project.org/package=stopp}, version 0.2.4. 

The research literature on spatial statistics provides a large body of techniques for analysing spatio-temporal point patterns, most of which are summarized in \cite{gonzalez:16}. Still, only a few of them have been implemented in software for general use. Some packages dealing with spatio-temporal point pattern exploratory analysis include \pkg{stpp} \citep{stpp,gabriel:rowlingson:diggle:2013}, \pkg{stppSim} \citep{stppSim}, \pkg{splancs} \citep{splancs}, and \pkg{stlnpp} \citep{stlnpp} whose main functionalities are summarised in Table \ref{tab:paks}. These include all the tools also provided by \pkg{stopp}.

\begin{table}[H]
	\centering
	\resizebox{\textwidth}{!}{
		\begin{tabular}{|l|cccccc|}
			\hline
			& \textbf{Simulations} & \textbf{Exploratory analysis} & \textbf{Model fitting} & \textbf{Diagnostics} & \textbf{Linear networks} & \textbf{Local analyses} \\
			\hline
			\pkg{stpp}  & \cmark  & \cmark  &    &    &    &   \cmark  \\
			\pkg{stppSim}  & \cmark  &    &    &    &    &    \\
			\pkg{splancs}   &    & \cmark  & \cmark  &    &    &    \\
			\pkg{stlnpp}  & \cmark  & \cmark  &    &    & \cmark   &    \\
			\hline
	\end{tabular}} 
	\label{tab:paks}
	\caption{List of \proglang{R} packages for spatio-temporal point processes and their main functionalities.}
\end{table}

While \pkg{stpp} allows for the simulation of Poisson, inhibitive and clustered patterns, the \pkg{stppSim} package generates artificial spatio-temporal point patterns through the integration of microsimulation and agent-based models. 
Moreover, \pkg{splancs} fosters many tools for the analysis of both spatial and spatio-temporal point patterns, including three-dimensional kernel estimation, Monte-Carlo tests of space-time clustering, and the estimation of homogeneous spatial and temporal $K$-functions. Regarding model fitting functions, it is only possible to fit the Diggle-Rowlingson Raised Incidence Model.
Moving to spatio-temporal point patterns on linear networks, the package \pkg{stlnpp} provides tools to visualise and analyse such patterns, implementing network-tailored kernel densities and first- and second-order summary statistics.\\
Among those, \pkg{stpp} stands out as the most comprehensive spatio-temporal point process devoted package, furnishing statistical tools for analyzing the global and local second-order properties of spatio-temporal point processes, including estimators of the space-time inhomogeneous $K$-function and pair correlation function.
All in all, none of the spatio-temporal point process packages allows for the diagnostics of a general fitted model.

Specifically, methods for fitting both separable and non-separable spatio-temporal point process models have emerged in many disciplines, including epidemiology \citep{jalilian2021hierarchical,briz2023mechanistic,schoenberg2023estimating}, seismicity \citep{xiong2023setas,adelfio:chiodi:15_spatial, siino:mateu:adelfio:16} and fire mapping \citep{raeisi2021spatio} in the classical Euclidean space, and GPS data \citep{dangelo2021inhomogeneous},  crimes \citep{dangelo2021self}, and traffic accidents \citep{kalair2021non,chaudhuri2023spatio,gilardi2024nonseparable,alaimo2024semi} in the context of linear networks. Some also included variables external to the point pattern under analysis as spatio-temporal covariates assumed to influence the occurrence of points \citep{adelfio2020including}. 
However, most of these methods were very specific to the chosen model, and there are no software implementations of sufficient generality to fit realistic models to a real dataset.
Packages dealing with spatio-temporal point process model fitting include \pkg{etasFLP} \citep{etasFLP,chiodi2017mixed,adelfio2020including}, mainly devoted to the estimation of the components of an ETAS (Epidemic Type Aftershock Sequence) model for earthquake description with the non-parametric background seismicity  estimated through FLP (Forward Likelihood Predictive), \pkg{ETAS} \citep{ETAS0,ETAS} which fits the space-time ETAS model to earthquake catalogs using a stochastic ``declustering'' approach, and 
\pkg{stelfi} \citep{stelfi}, which allows for the fitting of spatio-temporal self-exciting models and LGCPs (log-Gaussian Cox processes).
Another worth-to-mention package that implements routines to simulate and fit LGCPs include \pkg{lgcp} \citep{taylor:davies:barry:15}, which allows the fitting using methods
of the moments and Bayesian inference for spatial, spatio-temporal,
multivariate and aggregated point processes. 
This package, however, does not handle for non-separable (and anisotropic)
correlation structures of the covariance structure of the GRF (Gaussian Random Field). 
Turning to the context of the most simple spatio-temporal Poisson point processes, only the package \pkg{ppgam} \citep{ppgam,wood2017generalized} allows for the fitting of this kind of processes, but restricting the possibility to Generalized Additive Models, excluding more simple models like homogeneous and inhomogeneous Poisson process. Finally, playing an important role in the \proglang{R} spatial statistics community outside the CRAN, the \proglang{R-INLA}  package \citep{rue2009approximate} allows LGCP estimation within the framework of Bayesian inference for latent Gaussian models.

All the aforementioned packages leave no doubt about the widespread usage of spatio-temporal point process theory and its application by the spatial statistics community working with spatio-temporal data. However, as noted, none of those packages allows for a complete analysis of real datasets, including exploratory analysis, model fitting, and diagnostics. In particular, a considerable lack is the possibility of fitting spatio-temporal models permitting the inclusion of the dependence on external covariates. \\
The main contribution of the \pkg{stopp} package is the collection of standard tools for a complete analysis of a spatio-temporal point pattern while also fostering functions for more detailed issues. Among the latter, we highlight some spatio-temporal local tools, which are becoming more and more used in real spatio-temporal data analysis. The \pkg{stopp} package further allows for the integration with the previously mentioned packages by only requesting the estimated intensity to be diagnosed.

One of the main contributions of \pkg{stopp} is  embodied in the \proglang{stppm} function, which provides the first choice in \proglang{R} to fit general spatio-temporal Poisson point process models. These models include both homogeneous and inhomogeneous processes, with options for parametric and non-parametric specifications of coordinates, external covariates, and multitype cases.
This is achieved following a cubature scheme \citep{d2023locally,d2024preprint}, which extends \cite{berman1992approximating}'s and \cite{baddeley2014logistic}'s algorithm from the purely spatial to the spatio-temporal context.

Another important contribution of \pkg{stopp} lies in the second-order based diagnostic techniques, which only utilize fitted intensities, making them applicable to any fitted model (whether Euclidean or network-based), even to those beyond the scope of \pkg{stopp}. This versatility is a significant strength of \pkg{stopp} and enhances the linkage to other point process packages.
As far as we are aware, there is currently no software implementation of any technique for fitting spatio-temporal point process models at the level of generality and flexibility that we propose. This is only achieved by \pkg{spatstat} \citep{spatstat} in the purely spatial point process framework.

\pkg{stopp} also provides codes related to methods and models for analysing complex spatio-temporal point processes proposed in the papers \cite{siino2018joint,siino2018testing,adelfio2020some,dangelo2021assessing,dangelo2021local,d2023locally}. 
A particular focus is given to both first-order and second-order \textit{local} characteristics. Regarding first-order estimation, \pkg{stopp} allows for the estimation of both local spatio-temporal Poisson and local log-Gaussian Cox processes (LGCP) models, that is, with spatio-temporal varying parameters.
As previously mentioned, an \proglang{R} package that implements routines to fit spatio-temporal LGCPs is \pkg{lgcp}, where the \textit{minimum contrast} method is used to estimate parameters assuming a separable structure of the covariance of the Gaussian random field (GRF). In addition, \pkg{stopp} also handles non-separable correlation structures of the covariance structure of the GRF by means of the \textit{joint minimum contrast} procedure \citep{siino2018joint}, with the further advantage of giving the possibility of estimating both (or either) first-order and second-order parameters locally \citep{d2023locally}.

The level of generality achieved by \pkg{stopp} is due to the integration with other well-established point processes \proglang{R} packages.
The main dependencies of the \pkg{stopp}  package are indeed \pkg{spatstat}, \pkg{stpp}, and  \pkg{stlnpp}.
We exploit many functions from \pkg{spatstat} when purely spatial tools are needed while performing spatio-temporal analyses.
Furthermore, we rely on \pkg{stpp}'s both global and local $K$-functions and pcfs estimators, 
to perform diagnostics based on second-order summary statistics \citep{gabriel2009second,adelfio2020some}. 
From \pkg{stlnpp}, we borrow the linear networks estimators counterparts \citep{moradi2020first}. 

The ambitious aim of this package is to contribute to the existing literature by gathering many of the most widespread methods for the analysis of spatio-temporal point processes into a unique package, which is intended to host many further extensions.

The outline of the paper conceptually follows the package structure, illustrated in Table 2.

First, in Section \ref{sec:main}, we introduce the main classes of objects for handling spatio-temporal point pattern objects. Some available datasets are introduced in Section \ref{sec:data}. 
Then, we present some novel functions to simulate specific classes of point processes in Section \ref{sec:sim}.
We then move to Section \ref{sec:test} with exploratory analysis carried out through the Local Indicators of Spatio-Temporal Association (LISTA) functions on linear networks, newly available in \proglang{R}. In the same exploratory context, we illustrate the function to perform a local test for assessing the presence of local differences in two point patterns.  
Then, in Section \ref{sec:models}, a large body of functions available for fitting models is presented, including the general Poisson model, which includes both homogeneous or inhomogeneous specification of the first-order intensity function that can depend on semiparametric effects of both coordinates or external covariates. The multitype point process is also available. There is also the possibility of fitting a separable Poisson process model on either the Euclidean space and networks, and LGCPs. Moreover, we illustrate some functions to fit local models, including the generic Poisson process and LGCPs.
Finally, methods to perform global and local diagnostics on both models for point patterns on planar and linear network spaces are presented in Section \ref{sec:diag}. 
The paper ends with some future developments in Section \ref{sec:concl}.

\newpage

\begin{table}[H]
	\centering
	\resizebox{\textwidth}{!}{\begin{tabular}{|ll|}
			\bottomrule
			\multicolumn{2}{|c|}{\textbf{Data types}} \\
			\proglang{stcov()} & Create and interpolate spatio-temporal covariates on a regular grid\\
			\proglang{stp()} & Create \proglang{stp} and \proglang{stlp} objects for point patterns storage* \\
			\proglang{stpm()} & Create \proglang{stpm} and \proglang{stlpm} objects for marked point patterns storage*\\
			\midrule
			\multicolumn{2}{|c|}{\textbf{Datasets}} \\
			\proglang{chicagonet} & Rescaled roads of Chicago (Illinois, USA) \\
			\proglang{greececatalog} & Catalog of Greek earthquakes\\
			\proglang{valenciacrimes} & Crimes in Valencia in 2019\\
			\proglang{valencianet} & Roads of Valencia, Spain\\
			\midrule
			\multicolumn{2}{|c|}{\textbf{Simulations}} \\
			\proglang{rETASlp()} & Simulate a spatio-temporal ETAS process on a linear network\\
			\proglang{rETASp()} & Simulate a spatio-temporal ETAS process\\
			\proglang{rstlpp()} & Simulate spatio-temporal Poisson point patterns on a linear network\\
			\proglang{rstpp()} & Simulate spatio-temporal Poisson point patterns\\
			\midrule
			\multicolumn{2}{|c|}{\textbf{Eploratory analysis}} \\
			\proglang{localSTLginhom()} &Estimate the local inhomogeneous spatio-temporal pcfs on a linear network \\
			\proglang{localSTLKinhom()} & Estimate the local inhomogeneous spatio-temporal $K$-functions on a linear network\\
			\proglang{localtest()} & Perform the test of local structure for spatio-temporal point processes*\\
			\midrule
			\multicolumn{2}{|c|}{\textbf{Model fitting}} \\
			\proglang{locstppm()} & Fit a local spatio-temporal Poisson process  \\
			\proglang{sepstlppm()} & Fit a separable spatio-temporal Poisson process  on a linear network\\
			\proglang{sepstppm()} & Fit a separable spatio-temporal Poisson process \\
			\proglang{stlgcppm()} &Fit global or local spatio-temporal log-Gaussian Cox processes   \\
			\proglang{stppm()} & Fit a spatio-temporal Poisson process \\
			\midrule
			\multicolumn{2}{|c|}{\textbf{Diagnostics}} \\
			\proglang{globaldiag()} & Perform lobal diagnostics of a spatio-temporal point process models*\\
			\proglang{infl()} & Display outlying LISTA functions*\\
			\proglang{localdiag()} & Perform local diagnostics of spatio-temporal point process models*\\
			\bottomrule
	\end{tabular}}
	\label{tab:funs}
	\caption{List of functions in \pkg{stopp}, excluding S3 methods. The symbol * indicates the functions implemented to work both on point patterns in Euclidean spaces and linear networks.}
\end{table}

\newpage

\section{Data types}\label{sec:main}

\subsection{Spatio-temporal point patterns}

\begin{figure}[H]
	\centering
	\includegraphics[width=\textwidth]{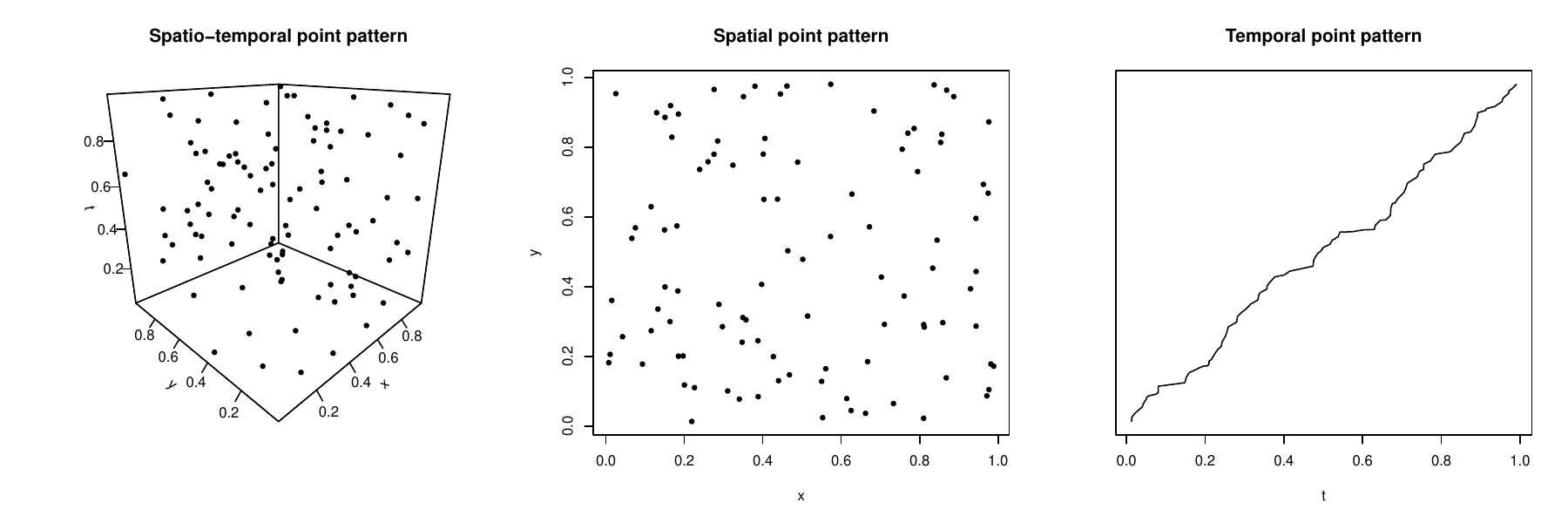}
	\caption{Output of the \proglang{plot.stp} function applied to a simulated spatio-temporal point pattern.}
	\label{fig:stp}
\end{figure}

\begin{figure}[H]
	\centering
	\includegraphics[width=\textwidth]{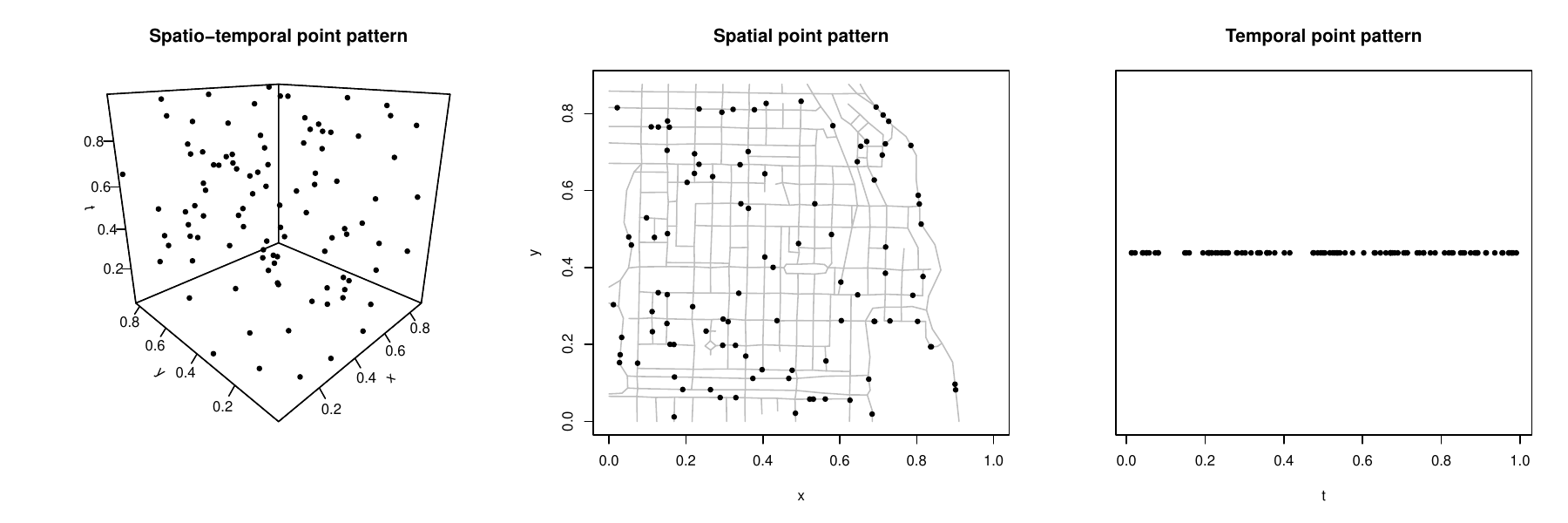}
	\caption{Output of the \proglang{plot.stlp} function applied to a simulated spatio-temporal point pattern on a linear network.}
	\label{fig:stpL}
\end{figure}

The \proglang{stp} function creates a \proglang{stp} object as a dataframe with three columns: \proglang{x}, \proglang{y}, and \proglang{t}.  If the linear network \proglang{L}, of class \proglang{linnet} of the \pkg{spatstat} package, is also provided, a \proglang{stlp} object is created instead.
This class of objects are equipped with the \proglang{print}, \proglang{summary}, and \proglang{plot} methods. 
The creation of these two types of objects comes as follows, with output plots provided in Figures \ref{fig:stp} and \ref{fig:stpL}, respectively.

\begin{Sinput}
R> install.packages("stopp")
R> library("stopp")  
R> set.seed(2)
R> df <- data.frame(runif(100), runif(100), runif(100))
R> stp1 <- stp(df)
R> stp1
\end{Sinput}

\begin{Soutput}
Spatio-temporal point pattern 
100 points 
Enclosing window: rectangle = [0.007109, 0.9889022] x [0.0136249, 0.9806] units
Time period: [0.013, 0.991]
\end{Soutput}

\begin{Sinput}
R> plot(stp1)
\end{Sinput}

The left and central panels produced by \proglang{plot.stp} and \proglang{plot.stlp} show the spatio-temporal and the purely spatial locations of the points. The right panel displays the cumulative sum of the temporal locations ordered in time. 
For this reason, the temporal cumulative plot of a homogeneous point pattern will be quadratic, and not linear as the intensity trend would be instead.
By setting the argument \proglang{tcum} equal to \proglang{FALSE}, the temporal pattern is displayed instead (Figure \ref{fig:stpL}), only advisable when dealing with few points.

\begin{Sinput}
R> set.seed(2)                       
R> df_net <- data.frame(runif(100, 0, 0.85), runif(100, 0, 0.85), runif(100))
R> stlp1 <- stp(df_net, L = chicagonet)
R> stlp1
\end{Sinput}

\begin{Soutput}
Spatio-temporal point pattern on a linear network 
100 points
Linear network with 338 vertices and 503 lines
Enclosing window: rectangle = [0, 0.9996963] x [0, 0.8763407] units 
(one unit = 1281.98625717162 feet)
Time period: [0.013, 0.991] 
\end{Soutput}
\begin{Sinput}
R> plot(stlp1, tcum = FALSE)
\end{Sinput}

\subsection{Marked point processes}

If additional variables are attached to the points of the pattern, it is possible to build a spatio-temporal marked point pattern as a \proglang{stpm} object (or \proglang{stlpm}, if occurred on a linear network).
For the multitype point process, we choose the same approach of continuous marks, that is, collecting all the points together in one point pattern and labelling each point by the type to which they belong. An advantage of this approach is that it is easy to deal with multitype point patterns with more than two types.

\begin{figure}[H]
	\centering
	\includegraphics[width=.8\textwidth]{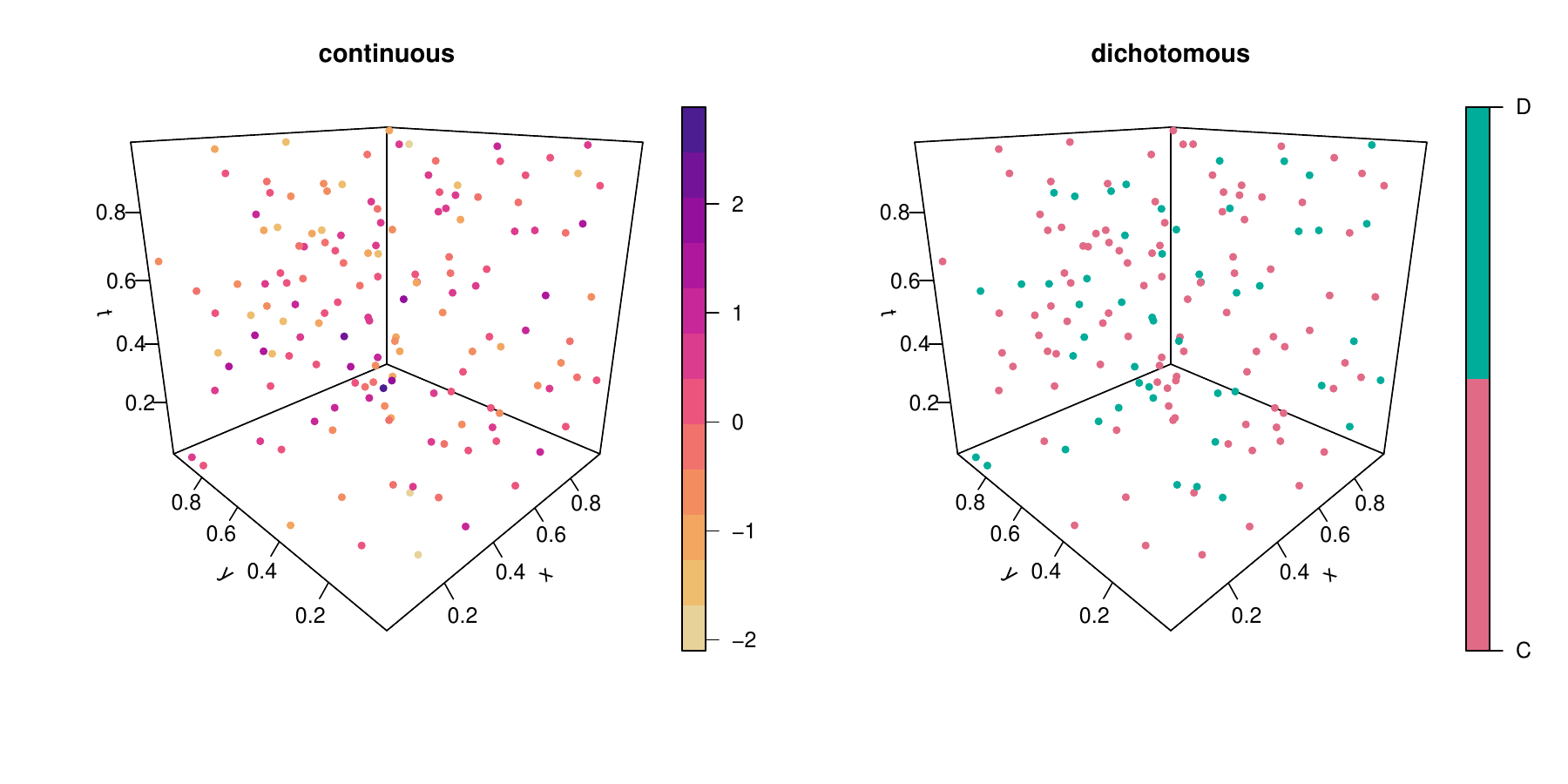}
	\caption{Output of the \proglang{plot.stpm} function applied to a simulated spatio-temporal point pattern marked by a continuous and a categorical mark.}
	\label{fig:mark2}
\end{figure}

Below is an example of a point pattern characterized by both a continuous mark and a categorical mark, rendering it a multitype point pattern, as shown in Figure~\ref{fig:mark2}.

\begin{Sinput}
R> set.seed(2)
R> dfA <- data.frame(x = runif(100), y = runif(100), t = runif(100), 
+    m1 = rnorm(100), m2 = rep(c("C"), times = 100))
R> dfB <- data.frame(x = runif(50), y = runif(50), t = runif(50), 
+    m1 = rnorm(25), m2 = rep(c("D"), times = 50))
R> stpm2 <- stpm(rbind(dfA, dfB), names = c("continuous", "dichotomous"))
R> plot(stpm2)
\end{Sinput}

\subsection{Spatio-temporal covariates}

The class \proglang{stcov} is reserved to be used for creating and interpolating potential spatio-temporal covariates, intended to be included in the \proglang{formula} of the main function of \pkg{stopp}: \proglang{stppm}. 

\begin{figure}[H]
	\centering
	\includegraphics[width=.4\textwidth]{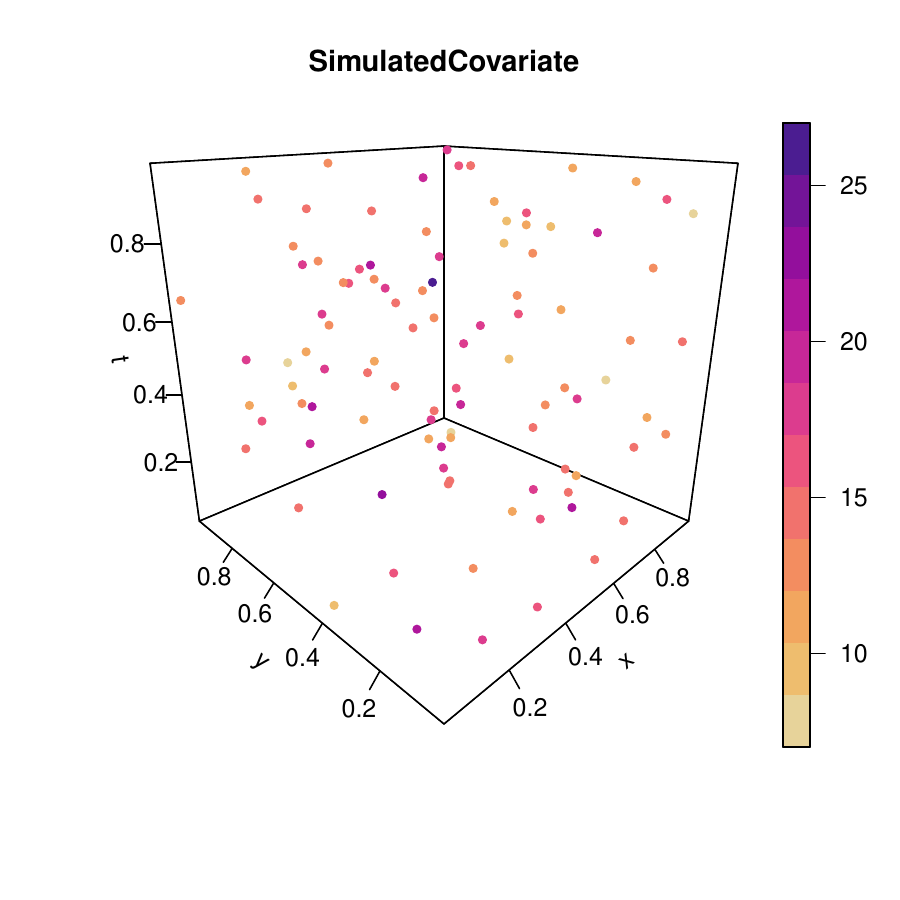}
	\includegraphics[width=.4\textwidth]{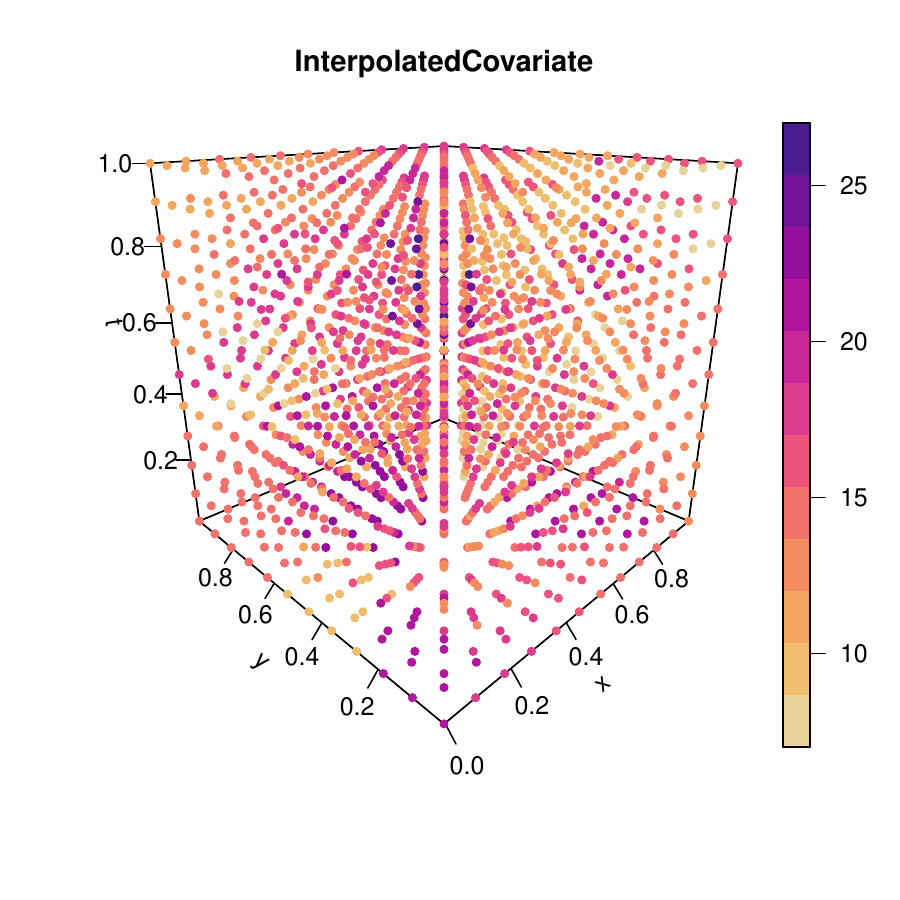}
	\caption{Simulated and interpolated covariate with the \proglang{stcov} function.}
	\label{fig:cov}
\end{figure}

Figure~\ref{fig:cov} displays an example of a simulated spatio-temporal covariate (on the left panel) and the interpolated covariate resulting from the application of the \proglang{stcov} function (right panel).

This preliminary procedure is a device to speed estimation in \proglang{stppm}. Indeed, since the covariate values must be known at every data and dummy point, an advisable approach is to use interpolation \citep{tarantino2024sis, d2024preprint}. We employ a spatial smoothing of the numeric values observed at the covariate locations
$
\hat{Z}(x,y,t) = \sum_{j=1}^J w_j(x,y,t) Z(x_j,y_j,t_j)/\sum_{j=1}^J w_j(x,y,t),
$
where $\hat{Z}(x,y,t)$ is the interpolated value at new location $(x,y,t)$, $J$ is the number of covariate locations, and $Z(x_j,y_j,t_j)$ is the covariate value at the observed location $(x_j,y_j,t_j)$.
Particularly, we set $w_j(x,y,t)= \big(\sqrt{(x-x_j)^2-(y-y_j)^2-(t-t_j)^2}\big)^{-p}$, meaning that we employ inverse-distance weighting \citep{shepard1968two}, where $p$ is the power of the Euclidean distance between $(x,y,t)$ and $(x_j,y_j,t_j)$. To avoid a different interpolation at each model fit, we, therefore, interpolate only once when employing the \proglang{stcov} function, making a very fine regular grid, and then just attribute to the data or dummy point the covariate value of the closest grid point in \proglang{stppm}.

\begin{Sinput}
R> set.seed(2)
R> df <- data.frame(runif(100), runif(100), runif(100), rpois(100, 15))
R> sim_cov <- stcov(df, interp = FALSE, names = "SimulatedCovariate")
R> interp_cov <- stcov(df, mult = 20, names = "InterpolatedCovariate")
R> plot(sim_cov)
R> plot(interp_cov)
\end{Sinput}

\section{Datasets}\label{sec:data}

\begin{figure}[H]
	\centering
	\includegraphics[width=\textwidth]{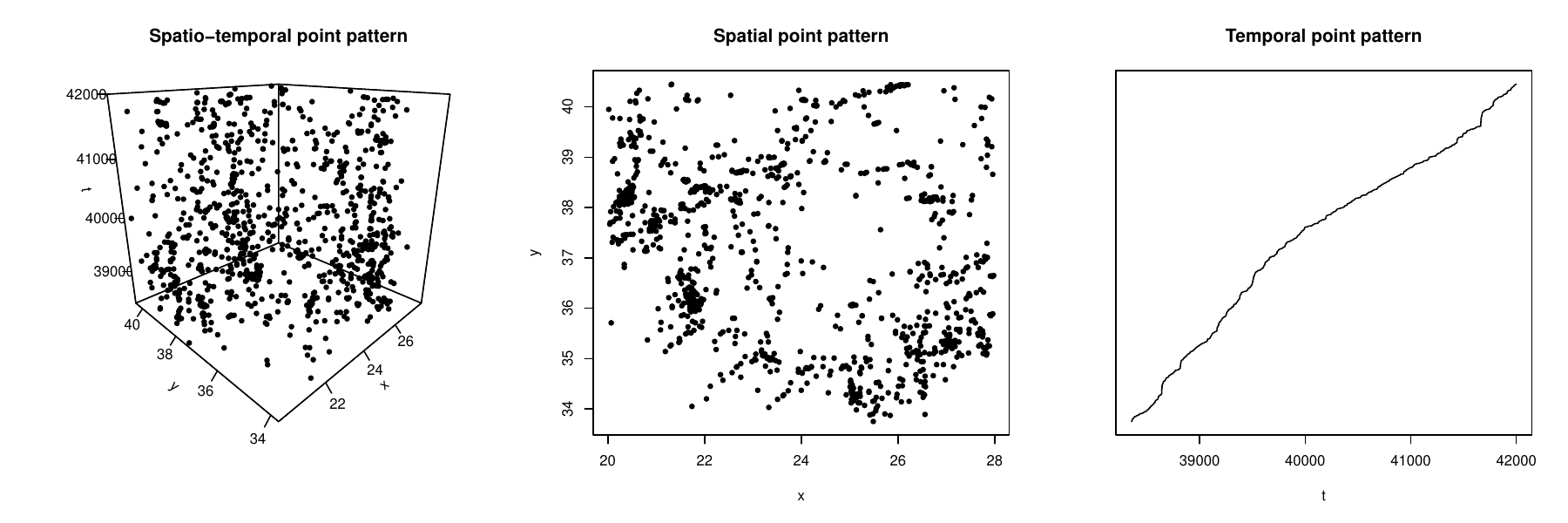}
	\caption{Plots of \proglang{greececatalog} data provided in \proglang{stp} format in the \pkg{stopp} package.}
	\label{fig:p2}
\end{figure}

The package is furnished with the \proglang{greececatalog} dataset\footnote{Data come from the Hellenic Unified Seismic Network (H.U.S.N.).} in the \proglang{stp} format containing the catalog of 
Greek earthquakes of magnitude at least 4.0 from 2005 to 2014 (Figure~\ref{fig:p2}). 

\begin{Sinput}
R> data("greececatalog", package = "stopp")
R> plot(greececatalog)
\end{Sinput}

\begin{figure}[H]
	\centering
	\includegraphics[width=\textwidth]{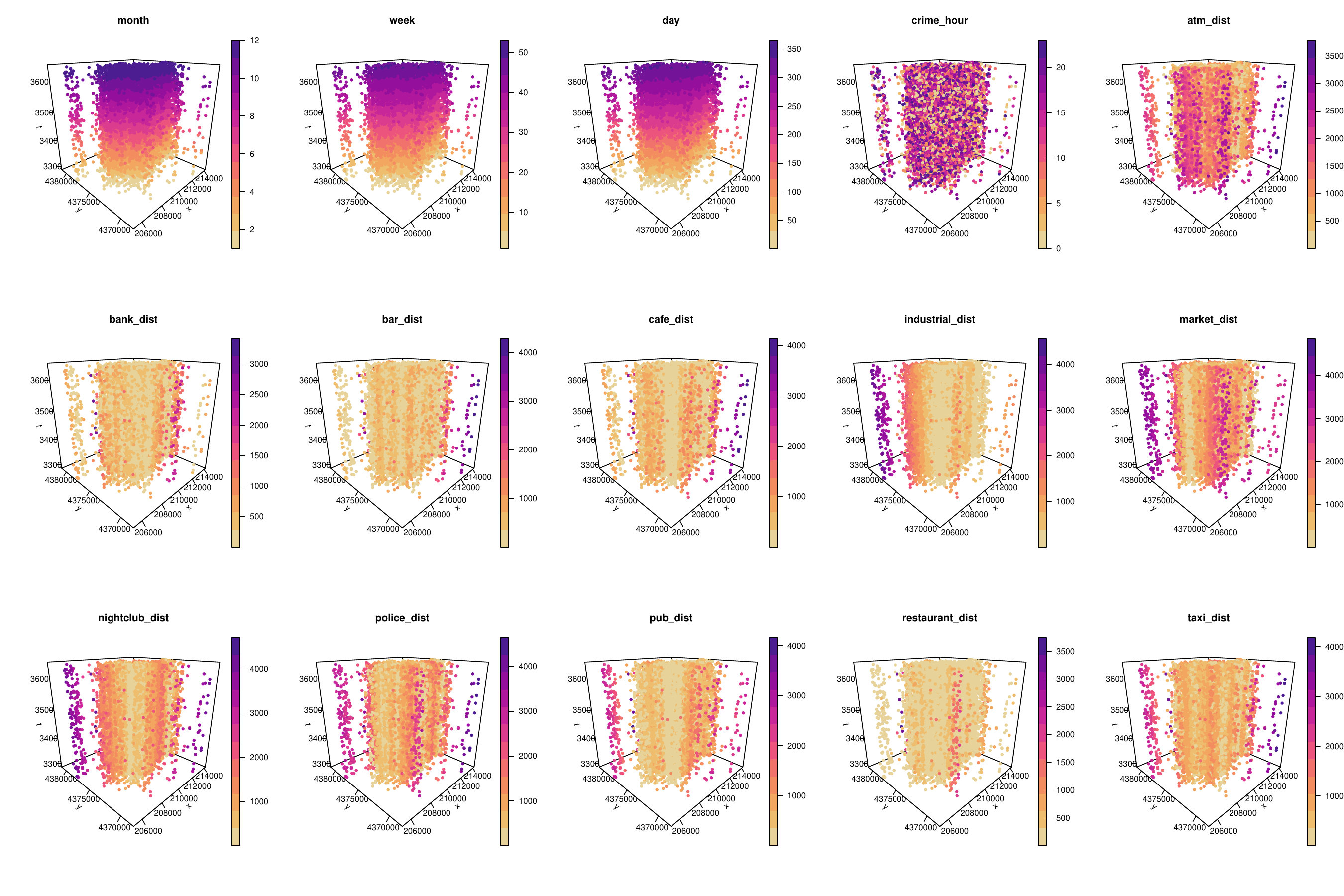}
	\caption{Plot of \proglang{valenciacrimes} data provided in \proglang{stpm} format in the \pkg{stopp} package.}
	\label{fig:valencia}
\end{figure}

\begin{figure}[H]
	\centering
	\includegraphics[width=.8\textwidth]{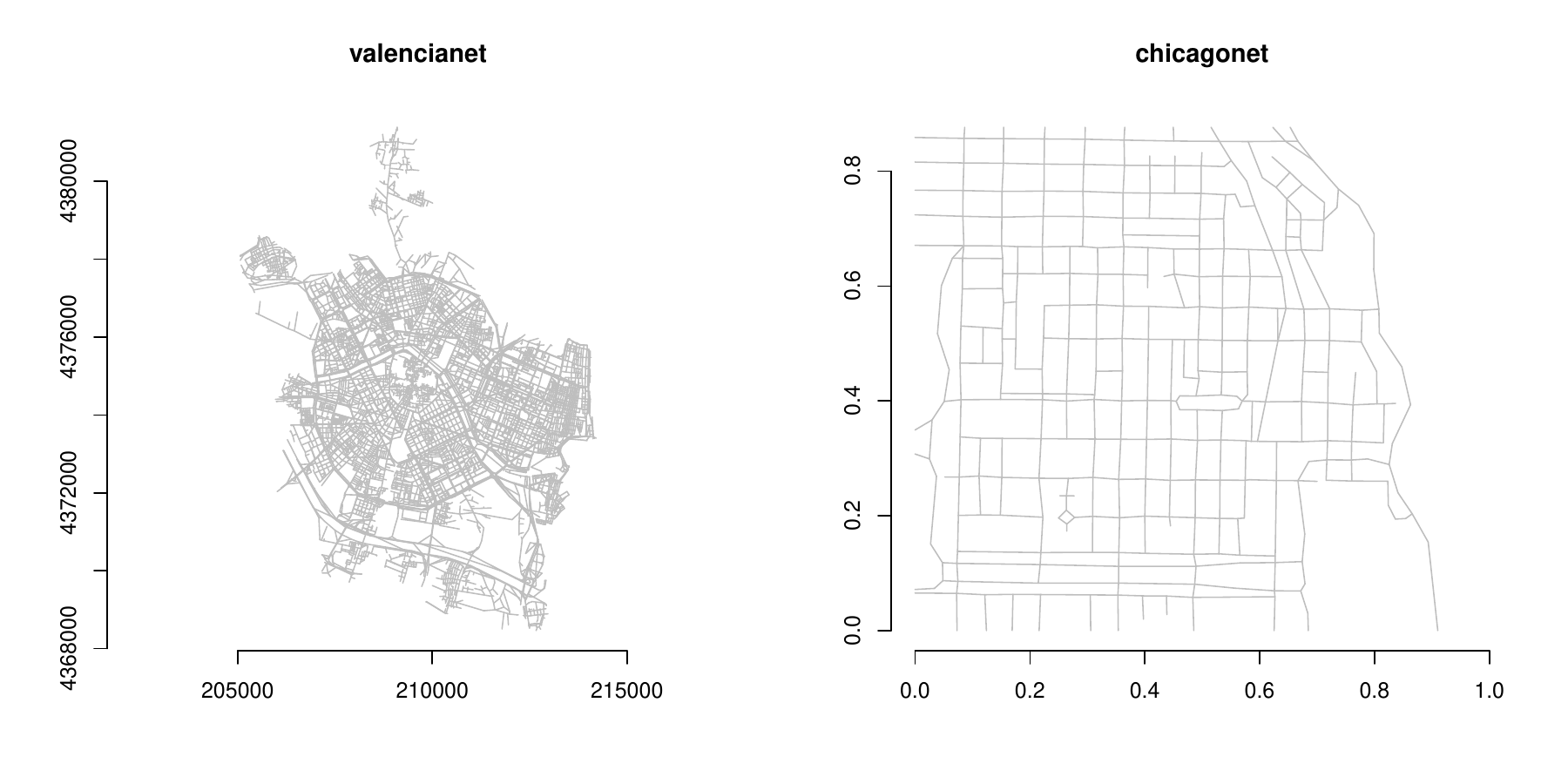}
	\caption{Plot of \proglang{valencianet} and \proglang{chicagonet} linear networks provided in the \pkg{stopp} package.}
	\label{fig:nets}
\end{figure}

A dataset of crimes that occurred in Valencia, Spain, in 2019 is also available in \proglang{stpm},
together with the linear network of class \proglang{linnet} of the Valencian roads, 
named  \proglang{valenciacrimes} (Figure~\ref{fig:valencia}),
and \proglang{valencianet} (right panel of Figure~\ref{fig:nets}), respectively.
The marks of this dataset include the month, week, day, and hour of crime occurrences, and many distances to the closest points of interest, which can be assumed to have influenced the occurrence of crimes.

\begin{Sinput}
R> data("valenciacrimes", package = "stopp")
R> plot(valenciacrimes)
R> data("chicagonet", package = "stopp")
R> data("valencianet", package = "stopp")
R> plot(chicagonet)
R> plot(valencianet)
\end{Sinput}

Finally, the linear network of class \proglang{linnet} of the roads of Chicago (Illinois, USA) close to the University of Chicago is also available (left panel of Figure~\ref{fig:nets}).
It represents the linear network of the Chicago dataset published and analysed in \cite{ang2012geometrically}. The window has been rescaled to be enclosed in a unit square.

\section{Simulations}\label{sec:sim}

Stochastic simulation of spatio-temporal point process models is another area where the richness of the theoretical literature contrasts with the scarcity of stable public domain software.

We contribute to the 
framework of simulating spatio-temporal point process models with novel designed functions. The first contribution is given by the possibility of simulating Poisson patterns as \proglang{stp} objects, with inhomogeneous intensity 
by means of the \proglang{rstpp} function, as follows.

\begin{Sinput}
R> rstpp(lambda = 500)
R> rstpp(lambda = function(x, y, t, a) {exp(a[1] + a[2] * x)}, par = c(2, 6))
\end{Sinput}

The above code simulates two spatio-temporal point patterns. The first one follows the homogeneous intensity $\lambda(x,y,t) = 500$, while the second one is generated from the inhomogeneous intensity $\lambda(x,y,t) = \exp(2 + 6 x)$. In the former case, the simulated pattern will be completely random, while the second one will show a trend increasing along the x-coordinate. 

The \proglang{rstlpp} function creates a \proglang{stlp} object instead, simulating a spatio-temporal Poisson point pattern 
on a linear network.

\begin{figure}[H]
	\centering
	\includegraphics[width=.4\textwidth]{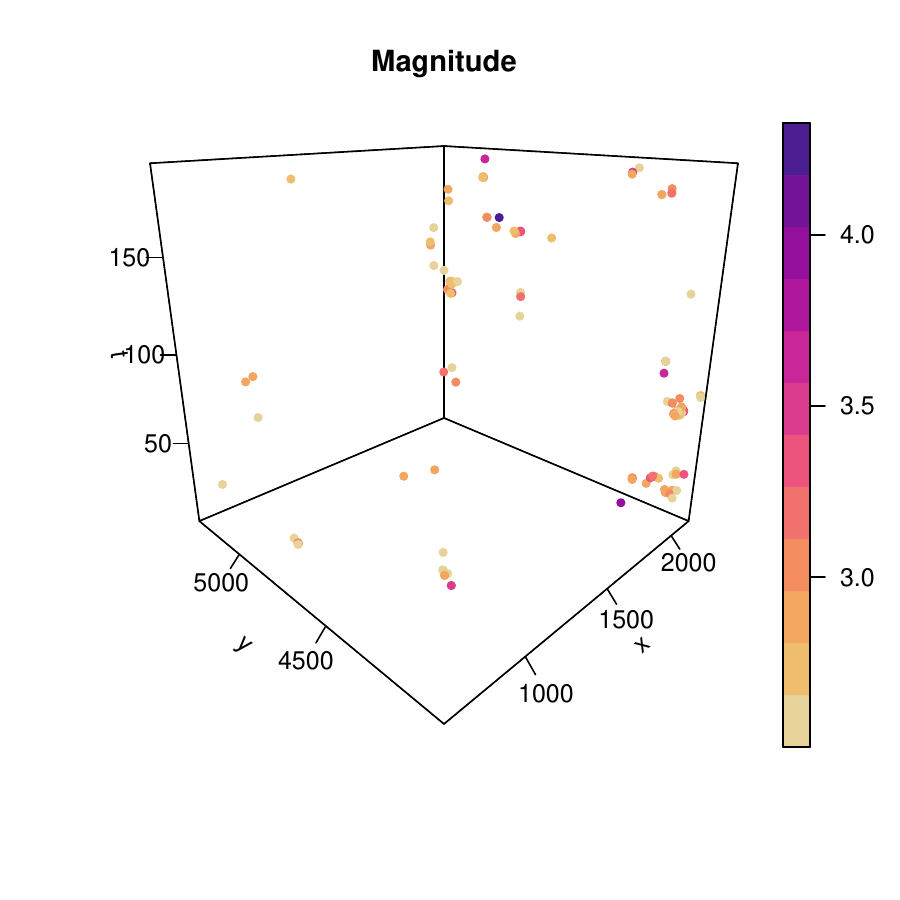}
	\caption{Plot of a spatio-temporal ETAS point pattern simulated by the \proglang{rETASp} function.}
	\label{fig:etas}
\end{figure}

Then, \proglang{rETASp} simulates a spatio-temporal point pattern following an Epidemic Type Aftershock Sequence (ETAS) process as in \cite{adelfio2020including}. Figure~\ref{fig:etas} shows an example.

\begin{Sinput}
R> set.seed(95)
R> X <- rETASp(c(0.1293688525, 0.003696, 0.013362, 1.2, 0.424466, 1.164793), 
+    betacov = 0.5, xmin = 600, xmax = 2200, ymin = 4000, ymax = 5300)
R> plot(X)
\end{Sinput}

Finally,  \proglang{rETAStlp} function creates a \proglang{stlp} object, simulating a spatio-temporal ETAS  process on a linear network. The simulation scheme in this case is adapted for the space location of events to be constrained on a linear network, being firstly introduced and employed for simulation studies by \cite{dangelo2021assessing}.

All the simulation functions are equipped with a \proglang{seed} argument, allowing to specify the seed for reproducing the same simulation.
Note that we have set specific seeds throughout the paper to ensure the reproducibility of the codes.

\section{Exploratory analysis}\label{sec:test}

The exploratory analysis tools of \pkg{stopp} build upon the Local Indicators of Spatio-Temporal Association (LISTA) functions, which are defined as a set of functions that are individually associated with each one of the points of the point pattern, and can provide information about the local behaviour of the pattern \citep{anselin:95,siino2018testing}.

\begin{figure}[H]
	\centering
	\includegraphics[width=\textwidth]{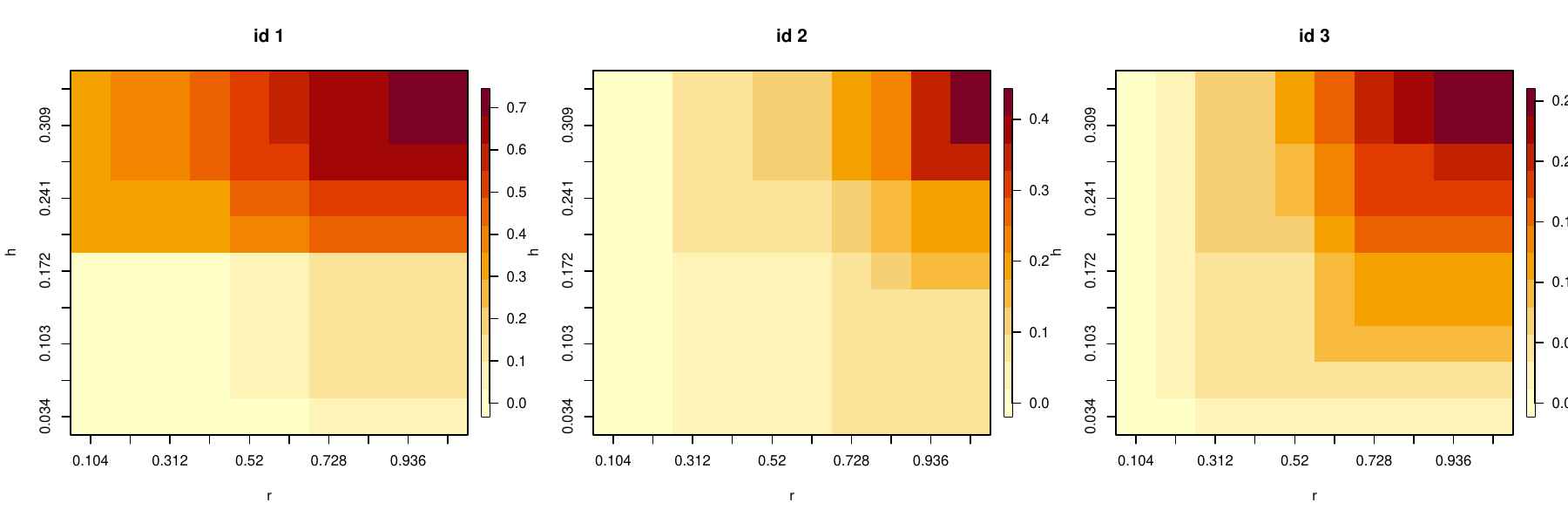}
	\caption{Output of the \proglang{plot.lista} function.}
	\label{fig:listas}
\end{figure}

In particular, the package implements the local spatio-temporal $K$-functions and pair correlation functions (pcfs) on linear networks, introduced in \cite{dangelo2021assessing}.
These are estimated by means of the function \proglang{localSTLKinhom} and \proglang{localSTLginhom}, respectively, and can be displayed through the \proglang{plot} function.
Since any of \proglang{localSTLKinhom} and \proglang{localSTLginhom} will produce a list of $K$-(or pcf)functions, one for each point in the observed point pattern, it is not possible to display them all together. Therefore, the argument  \proglang{id} is reserved for a vector for identifying which points to display the LISTA function of. Below is an example to display the local $K$-functions of the first three points stored in the \proglang{stp} object passed to the  \proglang{localSTLKinhom} function, as shown in Figure~\ref{fig:listas}. 

\begin{Sinput}
R> set.seed(2)
R> df_net <- data.frame(runif(25, 0, 0.85), runif(25, 0, 0.85), runif(25))
R> stlp1 <- stp(df_net, L = chicagonet)
R> lambda <- rep(diff(range(stlp1$df$x)) * diff(range(stlp1$df$y)) *
+    diff(range(stlp1$df$t)) / spatstat.geom::volume(stlp1$L), nrow(stlp1$df))
R> k <- localSTLKinhom(stlp1, lambda = lambda, normalize = TRUE)
R> plot(k, id = 1:3)
\end{Sinput}

\subsection{Local test}

\begin{figure}[H]
	\centering
	\includegraphics[width=\textwidth]{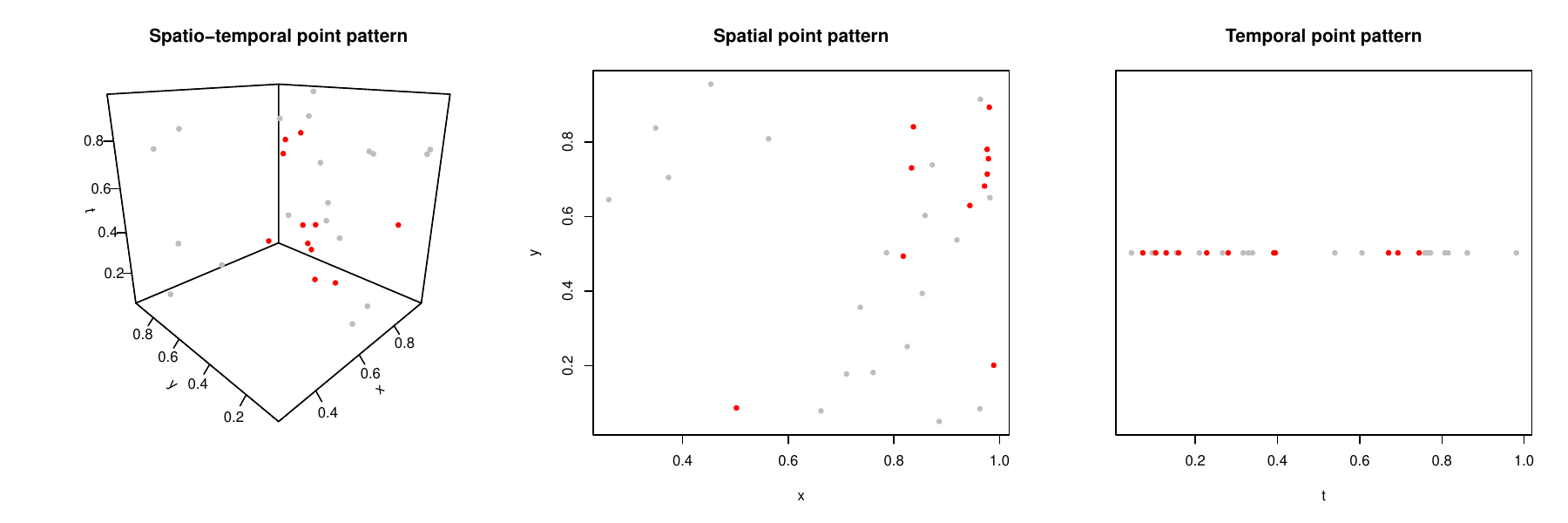}
	\caption{Output of the \proglang{plot.localtest} function.}
	\label{fig:p14}
\end{figure}

The function \proglang{localtest} performs the permutation test of the local structure of spatio-temporal point pattern data proposed in \cite{siino2018testing}.
The network counterpart is also implemented, following \cite{dangelo2021assessing}.
This test detects local differences in the second-order structure of two observed point patterns $\textbf{x}$ and $\textbf{z}$  
occurring in the same space-time region.
The test is  performed for spatio-temporal point patterns, as in
\cite{siino2018testing}, on two objects of class \proglang{stp}. The employed LISTA
functions $\hat{L}^{(i)}$  are the local $K$-functions introduced in 
\cite{adelfio2020some} and computed by the function 
\proglang{KLISTAhat} 
of the \pkg{stpp} package \citep{gabriel:rowlingson:diggle:2013}.
If \proglang{localtest} is applied to \proglang{stlp} objects, that is, on two spatio-temporal
point patterns observed on the same linear network \proglang{L}, 
the local $K$-functions
used are the ones proposed in \cite{dangelo2021assessing}, implemented
in the \proglang{localSTLKinhom} function of \pkg{stopp}. 
Details on the performance of the test are found in \cite{siino2018testing} and
\cite{dangelo2021assessing} for Euclidean and network spaces, respectively.
Alternative LISTA functions that can be employed to run the test are  \proglang{LISTAhat} of \pkg{stpp} and \proglang{localSTLginhom} of \pkg{stopp}, that is, the pcfs on Euclidean space and 
linear networks, respectively, fixing the argument \proglang{method = 'g'}.
The class of these objects is called \proglang{localtest}, and it is equipped with the methods \proglang{print}, \proglang{summary}, and \proglang{plot}, working as follows. In Figure~\ref{fig:p14}, an output example of the function \proglang{plot.localtest} is reported.
A background and an alternative patterns can be obtained, and the local test can be run as follows:
\begin{Sinput}
R> set.seed(2)
R> X <- rstpp(lambda = function(x, y, t, a) {exp(a[1] + a[2] * x)},
+    par = c(.005, 5))  
R> set.seed(2)
R> Z <- rstpp(lambda = 30)   
R> test <- localtest(X, Z, method = "K", k = 3)
R> test
\end{Sinput}

\begin{Soutput}
Test for local differences between two 
spatio-temporal point patterns 
--------------------------------------
Background pattern X: 30  
Alternative pattern Z: 25  

11 significant points at alpha = 0.05 
\end{Soutput}
\begin{Sinput}
R> plot(test)
\end{Sinput}

\section{Model fitting}\label{sec:models}

In this section, we outline the main functions to fit different specifications of inhomogeneous spatio-temporal Poisson process models.

\subsection{Homgeneous and inhomogeneous spatio-temporal Poisson point processes}

The primary fitting function of \pkg{stopp} is the function   \proglang{stppm}.
It fits a Poisson process model \citep{diggle2013statistical} to an observed spatio-temporal point pattern stored in a \proglang{stp} object,  assuming the template Poisson process model with a parametric first-order intensity function 
$$ \lambda(x,y, t; \boldsymbol{\theta}), \qquad
(x,y) \in W,\quad  t \in T , \quad \boldsymbol{\theta} \in \Theta,
$$
where $(x,y)$ and $t$ are the spatial and temporal coordinates in the spatial and temporal regions $W$ and $T$, and $\boldsymbol{\theta}$ are the parameters to be estimated.

For the homogeneous case, we can fit 
$$\lambda(x,y,t)=\lambda = \exp(\theta_0)$$
as follows:

\begin{Sinput}
R> set.seed(2)
R> ph <- rstpp(lambda = 200)
R> hom1 <- stppm(ph, formula = ~ 1, seed = 2)
R> hom1
\end{Sinput} 

\begin{Soutput}
Homogeneous Poisson process 
with Intensity: 202.093

Estimated coefficients:
(Intercept)
5.309
\end{Soutput}      

Therefore, the only mandatory arguments are the spatio-temporal point pattern \proglang{stp}, and the formula specifying the linear predictor to consider.
Note that the function \proglang{stppm} is also equipped with the argument \proglang{seed} since the generation of the dummy points depends on the \proglang{rstpp} function in turn. To make the code results reproducible, we set the seed in the examples illustrated with \proglang{stppm}, and in all the functions based on the generation of some dummy points.

In point process theory, it is common not to have available auxiliary covariates, so many point process models only resort to the Cartesian coordinates.

For the inhomogeneous case, we can simulate:

\begin{Sinput}
R> set.seed(2)
R> pin <- rstpp(lambda = function(x, y, t, a) {exp(a[1] + a[2] * x)}, 
+    par = c(2, 6))
\end{Sinput}

The following code fits a model with the following intensity specification
$$\lambda(x,y,t) = \exp(\theta_0 + \theta_1x)$$
estimating $\hat{{\theta}}_0 = 2.18$ and $\hat{{\theta}}_1 = 5.783$.

\begin{Sinput}
R> inh1 <- stppm(pin, formula = ~ x, seed = 2)
R> inh1
\end{Sinput}

\begin{Soutput}
Inhomogeneous Poisson process 
with Trend: ~x

Estimated coefficients: 
(Intercept)           x 
2.180       5.783
\end{Soutput}

Estimation is performed by fitting a Generalized Linear Mixed Model \citep{breslow1993approximate}, in which the linear predictor can contain random effects in addition to the usual fixed effects, employing a spatio-temporal cubature scheme \citep{d2023locally,d2024preprint}. The \proglang{stppm} function has an argument \proglang{method} which selects the parameter estimation technique.
Another option is \proglang{method='lsr'} representing the spatio-temporal extension of logistic spatial regression \citep{baddeley2014logistic}.
The choice of the \proglang{gam} function of the \pkg{mgcv} package \citep{wood2017generalized} is due to the possibility of including both smooth terms of the covariates (typical in point process theory for the spatio-temporal coordinates) and random effects. The latter comes in aid when wishing to fit a multitype point pattern, where basically each type of the categorical mark believed to represent the type will have its own set of fitted parameters \citep{d2024preprint}.

For instance, the following code fits an inhomogeneous Poisson point process of the form
$$\lambda(x,y,t) = \exp(f(x,y))$$
with $f(\cdot)$ a non-parametric function for the spatial coordinates estimated through thin plate regression splines \citep{wood2003thin} with 30 knots.

\begin{figure}[H]
	\centering
	\includegraphics[width=.8\textwidth]{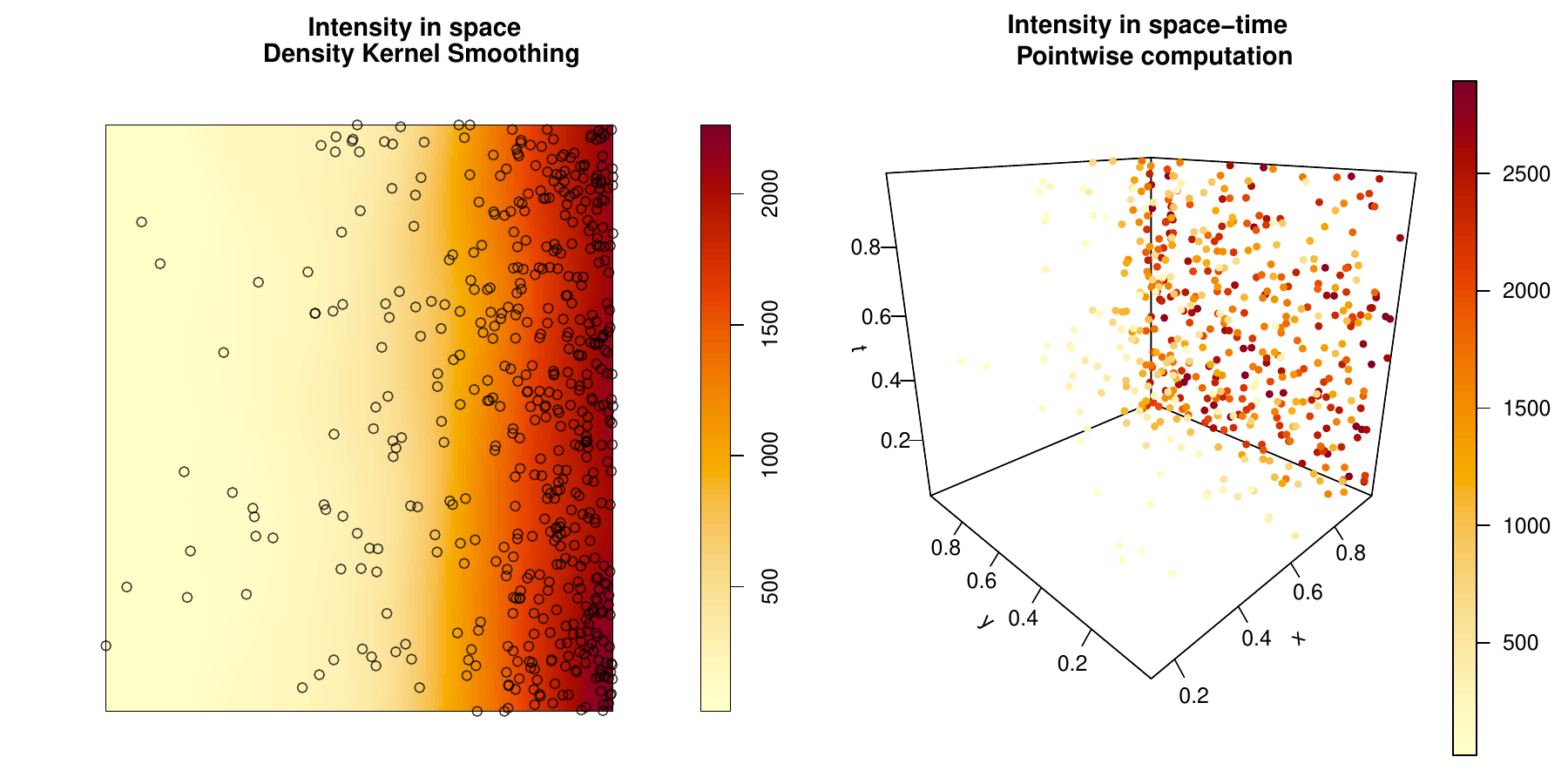}
	\caption{Output of the \proglang{plot.stppm} function applied to a fitted non-parametric model.}
	\label{fig:nonpar}
\end{figure}

Figure~\ref{fig:nonpar} shows the estimated intensity in space (left panel) and in space and time (right panel).

\begin{Sinput}
R> inh2 <- stppm(pin, formula = ~ s(x, y, bs = "tp",  k = 30), seed = 2)
R> plot(inh2)
\end{Sinput}

\subsection{Spatio-temporal Poisson point processes with dependence on external covariates}

Another peculiar capability in \pkg{stopp} is the possibility of fitting Poisson point process models with a first-order intensity function depending on external spatio-temporal covariates as
$$ \lambda(x,y, t; \boldsymbol{\theta}) = \exp(\boldsymbol{\theta}^{\top}\textbf{Z}(x,y,t)),
$$
where $\textbf{Z}(x,y, t)=\{Z_1(x,y, t), \ldots, Z_p(x,y, t)\}$ are $p$ known spatio-temporal covariate functions, and $\boldsymbol{\theta}$ their associated unknown parameters to estimate.

It is very uncommon to have the covariate values observed at the point pattern locations. Nevertheless, their values must be known at points and some other locations in the analysed region for inferential purposes. This is achieved by preliminary interpolating the covariate values through the \proglang{stcov} function, as shown in the example below.

Let's first simulate some covariates.

\begin{Sinput}
R> set.seed(2)
R> df1 <- data.frame(runif(100), runif(100), runif(100), rpois(100, 15))
R> df2 <- data.frame(runif(100), runif(100), runif(100), rpois(100, 15))
\end{Sinput}

Next, it is advisable to interpolate them along a finer and more regular grid with \proglang{stcov}, which will return a \proglang{stcov} object. 

\begin{Sinput}
R> obj1 <- stcov(df1, names = "cov1")
R> obj2 <- stcov(df2, names = "cov2")
\end{Sinput}

Then, we have to store all of the covariates into a unique list.

\begin{Sinput}
R> covariates <- list(cov1 = obj1, cov2 = obj2)
\end{Sinput}

Note that this is necessary because, often, the covariate's sites are not the same among different covariates. 
To then fit a spatio-temporal Poisson point process model depending on a spatial coordinate and a spatio-temporal covariate, 
such as
$$\lambda(x,y,t) = \exp(\theta_0 + \theta_1x + \theta_2 \texttt{cov2}(x,y,t)),$$
we have to input the list of \proglang{stcov} objects into the \proglang{covs} argument of \proglang{stppm} and specify \proglang{spatial.cov = TRUE}, as the following code illustrates.

\begin{Sinput}
R> inh3 <- stppm(pin, formula = ~ x + cov2, covs = covariates,  
+    spatial.cov = TRUE, seed = 2)
R> inh3
\end{Sinput}

\begin{Soutput}
Inhomogeneous Poisson process 
with Trend: ~x + cov2

Estimated coefficients: 
(Intercept)           x        cov2 
2.116       5.791       0.004
\end{Soutput}

\subsection{Multitype spatio-temporal Poisson point processes}

Finally, \proglang{stppm} offers the capability to fit multitype Poisson point process models. 

If the multitype point process has $m = 1,2, \ldots ,M$ types, the (marginal) intensity is $$\lambda(x,y,t) = \sum_{m=1}^{M}\lambda(x,y,t,m) $$ where $\lambda(x,y,t,m)$ is the intensity function for locations $(x,y,t)$ and mark type $m$.

\begin{figure}[H]
	\centering
	\includegraphics[width=.4\textwidth]{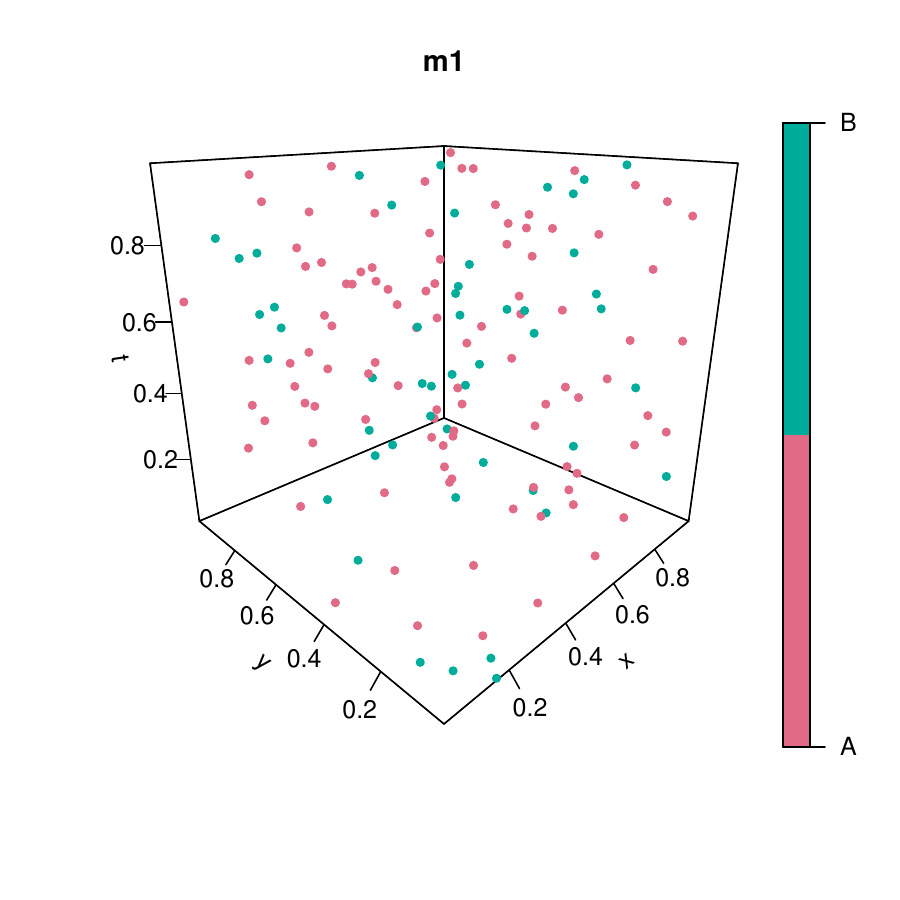}
	\caption{Plot of a simulated multitype point pattern with only two types of points.}
	\label{fig:multi}
\end{figure}

As an example, the following codes simulate a multitype point pattern with points belonging to two different types, named A and B, with 100 and 50 points each (Figure~\ref{fig:multi}).

\begin{Sinput}
R> set.seed(2)
R> dfA <- data.frame(x = runif(100), y = runif(100), t = runif(100), 
+    m1 = rep(c("A"), times = 100))
R> dfB <- data.frame(x = runif(50), y = runif(50), t = runif(50), 
+    m1 = rep(c("B"), each = 50))
R> stpm1 <- stpm(rbind(dfA, dfB))
R> plot(stpm1)
\end{Sinput}

To fit a multitype Poisson point process model, therefore, an object of \proglang{stpm}, with a categorical mark, must be provided to \proglang{stppm}. The multitype model is fitted by setting \proglang{marked = TRUE}, and by calling the mark with a formula like \proglang{s(mark, bs = "re")}, exactly following the random effects specifications of the \proglang{gam} function.
In brief, this is because multitype point process fitting is based on a cubature scheme replicated for each mark type.
For instance, the following code fits a multitype Poisson process model with inhomogeneous intensity depending on the x-coordinate and a random intercept ${\theta}_{0m}$, as follows
$$\lambda(x,y,t) = \exp(\theta_0 + {\theta}_{0m} + \theta_1x).$$

\begin{Sinput}
R> inh4 <- stppm(stpm1, formula = ~ x + s(m1, bs = "re"), marked = TRUE, 
+    seed = 2)
\end{Sinput}

\begin{figure}[H]
	\centering
	\includegraphics[width=.8\textwidth]{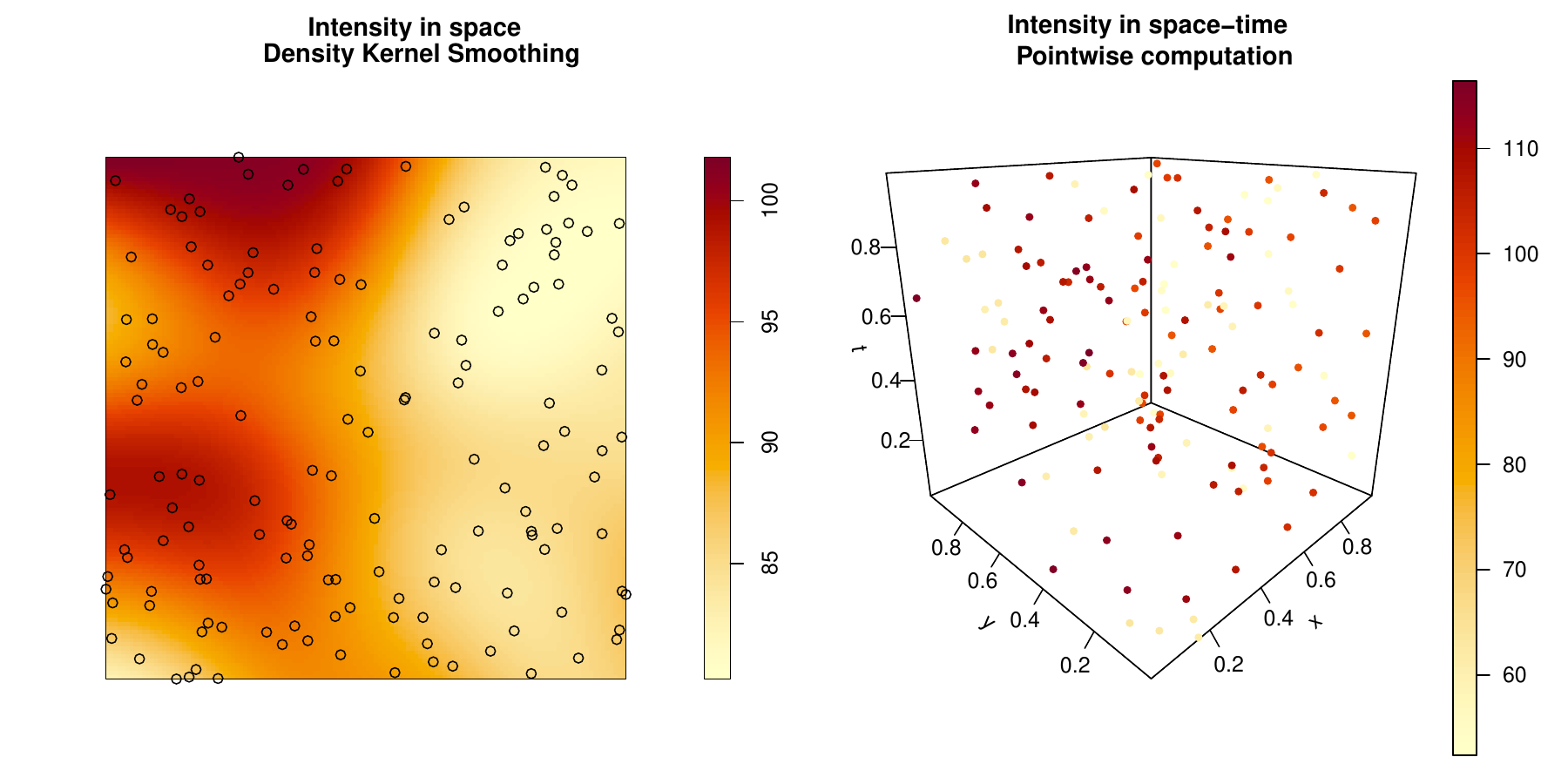}
	\caption{Output of the \proglang{plot.stppm} function applied to a fitted multitype model.}
	\label{fig:multimodel}
\end{figure}

In point process terms, this means that the average number of points will differ between the two types, but the x-coordinate is believed to have a common effect on the intensities of the two subpatterns.
The right panel of Figure~\ref{fig:multimodel} clearly illustrates these results, showing a consistently low intensity for the points belonging to the subpattern with fewer points.

\begin{Sinput}
R> plot(inh4)
\end{Sinput}

Note that any combination of the presented model specifications is allowed. For instance, multitypes point processes can be fitted, with semi-parametric specifications of the first-order intensity, depending on both coordinates and external spatio-temporal covariates.

\subsection{Spatio-temporal Poisson point processes with separable intensity}

\begin{figure}[H]
	\centering
	\includegraphics[width=.8\textwidth]{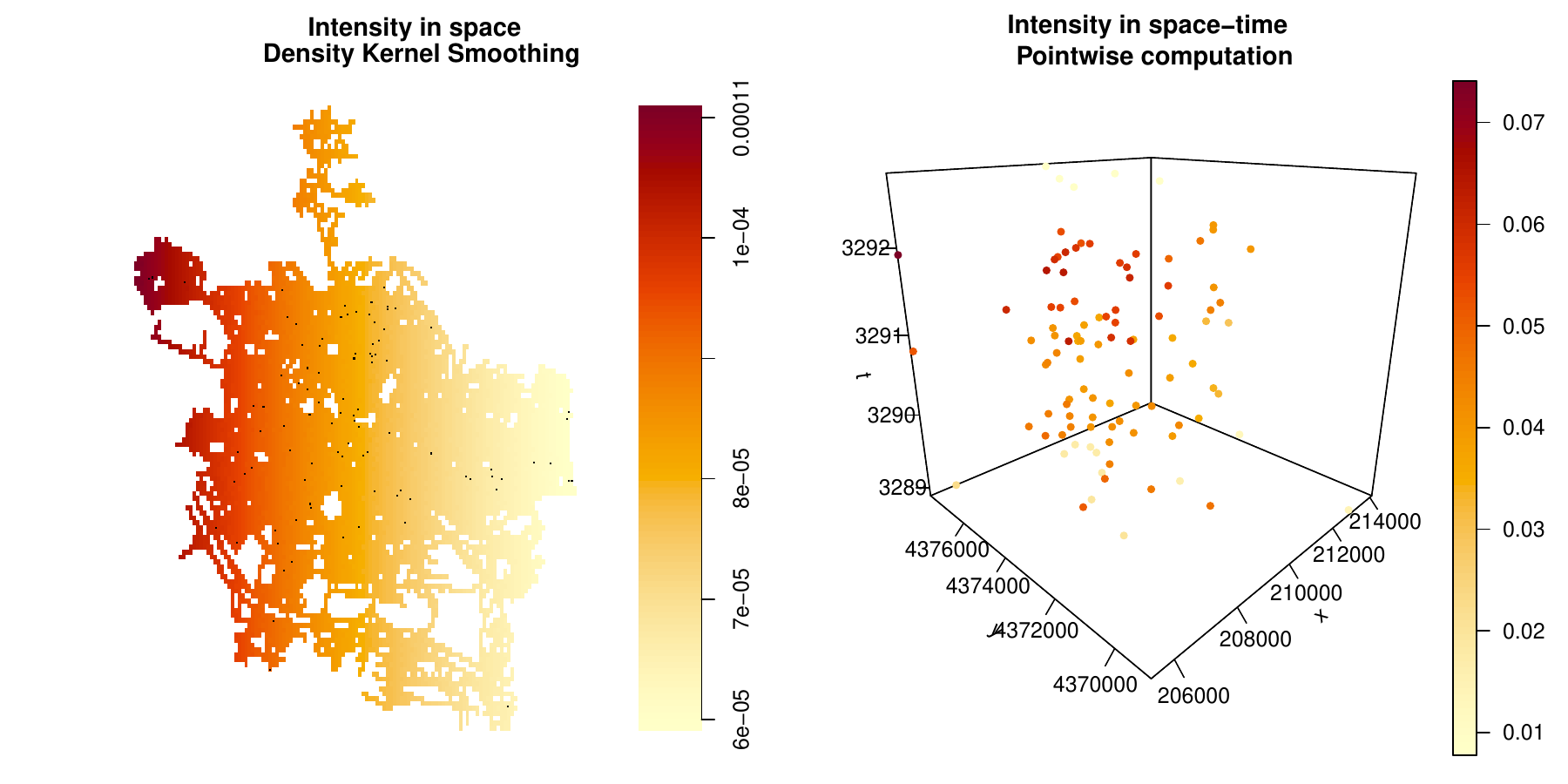}
	\caption{Plot of the separable model fitted produced by the \proglang{plot.sepstlppm} function.}
	\label{fig:sepL}
\end{figure}

The function \proglang{sepstppm} fits a separable parametric spatio-temporal Poisson process model \citep{diggle2013statistical} to point patterns observed on a subset of the Euclidean space, according to the following generic form
$$\lambda(x,y,t) = \lambda(x,y)\lambda(t),$$
where ${\lambda}(x,y)$ and ${\lambda}(t)$ are non-negative functions on $W$ and $T$, respectively.
This formulation can include a combination of a parametric spatial point pattern model, potentially depending on the spatial coordinates and/or spatial covariates, and a parametric log-linear model for the temporal component. 
The spatio-temporal intensity is therefore obtained by multiplying the spatial and temporal intensities fitted separately. This has the advantage of giving the possibility to include purely spatial and purely temporal covariates, denoted by $\textbf{Z}_{S}(x,y)$ and $\textbf{Z}_{T}(t)$, with the following general formulation
$$
\lambda(x,y,t) = \lambda(x,y)\lambda(t) = \exp(\theta_0 + \boldsymbol{\theta}_{S}^{\top}\textbf{Z}_{S}(x,y) + \boldsymbol{\theta}_{T}^{\top}\textbf{Z}_{T}(t)).$$
The function \proglang{sepstlppm} implements the network counterpart of the spatio-temporal Poisson point process with separable intensity and fully parametric specification.
Concerning linear network point patterns, only non-parametric estimators of the intensity function have been suggested in the literature \citep{mateu2020spatio,moradi2020first}.
The functions \proglang{plot.sepstppm} and \proglang{plot.sepstlppm} show the fitted intensities, displayed both in space and in space and time.
Next, we perform an example on a subset of the Valencia dataset, including the linear network in the inferential procedure. See Figure~\ref{fig:sepL} for the plot of the carried-out example.

\begin{Sinput}
R> crimesub <- stpm(valenciacrimes$df[101:200, ],
+    names = colnames(valenciacrimes$df)[-c(1:3)], L = valencianet)
R> mod1 <- sepstlppm(crimesub, spaceformula = ~x , timeformula = ~ day)
R> plot(mod1)
\end{Sinput}

\subsection{Spatio-temporal Poisson point processes with non-separable intensity}

When separability of the spatial and temporal component is not plausible for the data, a non-separable specification of the intensity function is more advisable. This is obtained through the \proglang{stppm} function when  including proper spatio-temporal covariates or specifying any kind of interaction between spatial and temporal variables.

As an example, the following code fits an inhomogeneous non-separable spatio-temporal Poisson model 
with dependence on the spatio-temporal coordinates and some of their polynomials and interactions specified as follows
$$\lambda(x,y,t) = \exp(\theta_0 + \theta_1x + \theta_2y + \theta_3t + \theta_4xy + \theta_5yt + \theta_6x^2 + \theta_7y^2 + \theta_8t^2 + \theta_9x^2y^2).$$

\begin{Sinput}
R> nonsepmod <- stppm(greececatalog, formula = ~ x + y + t + x:y + y:t +
+    I(x^2) + I(y^2) + I(t^2) + I(x^2):I(y^2), seed = 2)
\end{Sinput}

As any other model fitted through \proglang{stppm}, both the \proglang{print} and \proglang{summary} functions will return the estimated coefficients, and the \proglang{plot} function will display the estimated intensity in space and in space and time.

\begin{Sinput}
R> summary(nonsepmod)
\end{Sinput}
\begin{Soutput}
Inhomogeneous Poisson process 
with Trend: ~x + y + t + x:y + y:t + I(x^2) + I(y^2) + I(t^2) + I(x^2):I(y^2)

Estimated coefficients: 
(Intercept)            x            y            t       I(x^2)       I(y^2) 
-967.872       54.323       41.785       -0.007       -0.528       -0.343 
I(t^2)          x:y          y:t I(x^2):I(y^2) 
0.000        -1.481        0.000         0.000 
\end{Soutput}

\begin{Sinput}
R> plot(nonsepmod)
\end{Sinput}
\begin{figure}[H]
	\centering
	\includegraphics[width=.8\textwidth]{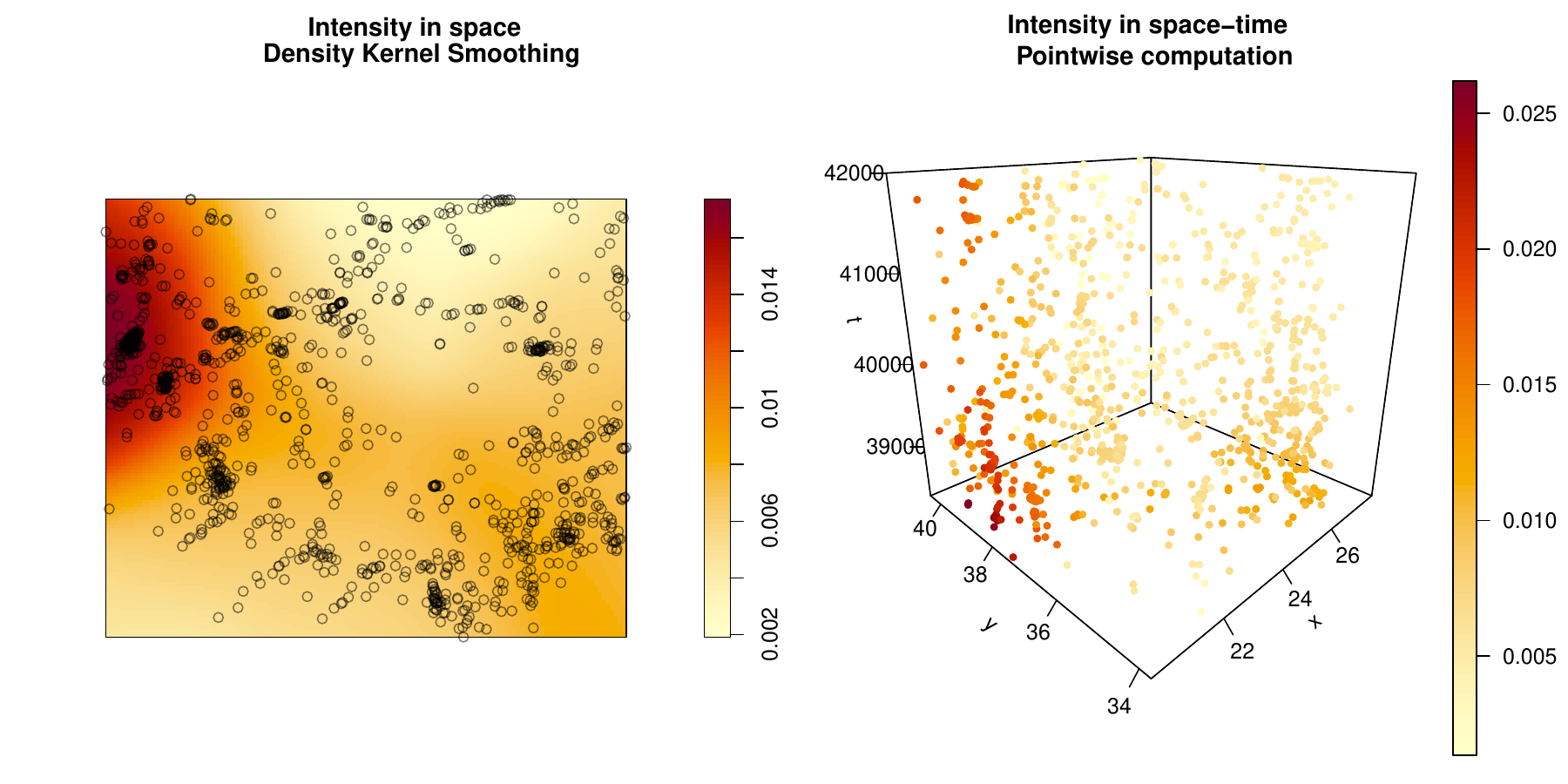}
	\caption{Plot of the non-separable model fitted produced by the \proglang{plot.stppm} function.}
	\label{fig:nonsepmod}
\end{figure}

Furthermore, since the model is fitted altogether employing a GLM, the significance of the parameters can be inspected by checking the \proglang{summary} of the \proglang{mod\_global} object returned by the \proglang{stppm} function.

\begin{Sinput}
R> summary(nonsepmod$mod_global)
\end{Sinput}

\begin{Soutput}
Family: poisson 
Link function: log 

Formula:
y_resp ~ x + y + t + x:y + y:t + I(x^2) + I(y^2) + I(t^2) + I(x^2):I(y^2)

Parametric coefficients:
Estimate Std. Error z value Pr(>|z|)    
(Intercept)   -9.679e+02  1.344e+02  -7.202 5.93e-13 ***
x              5.432e+01  6.687e+00   8.124 4.50e-16 ***
y              4.179e+01  4.597e+00   9.090  < 2e-16 ***
t             -7.294e-03  2.326e-03  -3.136 0.001712 ** 
I(x^2)        -5.282e-01  6.989e-02  -7.557 4.12e-14 ***
I(y^2)        -3.432e-01  3.328e-02 -10.311  < 2e-16 ***
I(t^2)         6.193e-08  2.869e-08   2.159 0.030867 *  
x:y           -1.481e+00  1.782e-01  -8.313  < 2e-16 ***
y:t            5.823e-05  1.622e-05   3.590 0.000331 ***
I(x^2):I(y^2)  3.920e-04  4.953e-05   7.914 2.49e-15 ***
---
Signif. codes:  0 ‘***’ 0.001 ‘**’ 0.01 ‘*’ 0.05 ‘.’ 0.1 ‘ ’ 1
\end{Soutput}

Note that even though the \proglang{mod\_global} object is of class \proglang{glm}, it is not advisable to rely on standard classical GLM tools, such as the AIC or the $R^2$, since they depend on the chosen structure of the cubature scheme, not explored here in detail.

\subsection{Log-Gaussian Cox processes}

The \proglang{stlgcppm} function estimates the covariance parameters of a spatio-temporal log-Gaussian Cox process (LGCP) \citep{diggle:moraga:13} with random intensity
$$\Lambda(x,y,t)=\lambda(x,y,t)\exp(S(x,y,t)),$$
following the joint minimum contrast procedure introduced in \cite{siino2018joint}.
LGCPs are hierarchical Poisson processes, where the dependence in the point pattern is modelled through the common latent Gaussian variable $S$ \citep{rue2009approximate}.
Here $S$ is a Gaussian process with $\mathbb{E}(S(x,y,t))=\mu=-0.5\sigma^2$ and so  $\mathbb{E}(\exp(S(x,y,t)))=1$ and with variance and covariance matrix $\sigma^2 \gamma(r,h)$ under the stationary assumption, with $\gamma(\cdot)$ the correlation function of the Guassian Random Field (GRF), and $r$ and $h$ some spatial and temporal distances. 

The covariances available are separable exponential, Gneiting, and Iaco-Cesare.
The function works by assuming a homogeneous first-order intensity as default.
Different inhomogeneous specifications of the first-order intensity function are implemented as well.

\begin{Sinput}
R> catsub <- stp(greececatalog$df[1:200, ])
R> lgcp1 <- stlgcppm(catsub, seed = 2)
\end{Sinput}

As a default, the package fits a LGCP model with a separable structure for the covariance function of the GRF \citep{brix2001spatiotemporal} that has exponential form for both the spatial and the temporal components,
\begin{equation*}
	\mathbb{C}(r,h)=\sigma^2\exp \bigg(\frac{-r}{\alpha}\bigg)\exp\bigg(\frac{-h}{\beta}\bigg),
\end{equation*}
where $\sigma^2$ is the variance parameter, $\alpha$ is the scale parameter for the spatial distance and $\beta$ is the scale parameter for the temporal one.

The \proglang{print} and \proglang{summary} functions give the main information on the fitted model. 

\begin{Sinput}
R> lgcp1
\end{Sinput}

\begin{Soutput}
Joint minimum contrast fit 
for a log-Gaussian Cox process with 
global first-order intensity and 
global second-order intensity 
--------------------------------------------------
Homogeneous Poisson process 
with Intensity: 0.00849

Estimated coefficients of the first-order intensity: 

(Intercept) 
-4.769 
--------------------------------------------------
Covariance function: separable 

Estimated coefficients of the second-order intensity: 

sigma  alpha   beta 
15.389  0.239 15.275 
--------------------------------------------------
Model fitted in 0.014 minutes 
\end{Soutput}

The \proglang{plot.sepstlppm} function shows the fitted intensity displayed 
both in space (by means of a density kernel smoothing) and in space and time, similar to what we have seen so far with the other classes of models. 

\subsection{Local models}

\subsubsection{Local spatio-temporal Poisson point processes}

The \proglang{locstppm} function fits a spatio-temporal local Poisson process model \citep{d2023locally} to an observed spatio-temporal
point pattern stored in a \proglang{stp} object, that is, a Poisson model with
a vector of parameters $\boldsymbol{\theta}_i \in \Theta$ for each point $(x_i,y_i,t_i)$.
In local likelihood estimation of Poisson processes \citep{loader1999bandwidth} the estimated intensity at $(x,y,t)$ is
taken to be the plug-in value $$\hat{\lambda}(x,y,t) = {\lambda(x,y,t; \hat{\boldsymbol{\theta}}(x,y,t))}$$ associated with the fitted parameter vector at $(x,y,t)$.

The \proglang{print} and \proglang{summary} functions will provide information of the estimated local parameters by means of the summary of their distributions.

\begin{Sinput}
R> set.seed(2)
R> inh <- rstpp(lambda = function(x, y, t, a) {exp(a[1] + a[2] * x)}, 
+    par = c(0.005, 5))
R> inh_local <- locstppm(inh, formula = ~ x, seed = 2)
R> inh_local
\end{Sinput}

\begin{Soutput}
Inhomogeneous Poisson process 
with Trend: ~x

Summary of estimated coefficients 
(Intercept)           x        
Min.   :0.3075   Min.   :2.803  
1st Qu.:0.9073   1st Qu.:3.652  
Median :1.4415   Median :4.264  
Mean   :1.4360   Mean   :4.291  
3rd Qu.:2.0157   3rd Qu.:4.975  
Max.   :2.7504   Max.   :5.637 
\end{Soutput}

\begin{figure}[H]
	\centering
	\includegraphics[width=.8\textwidth]{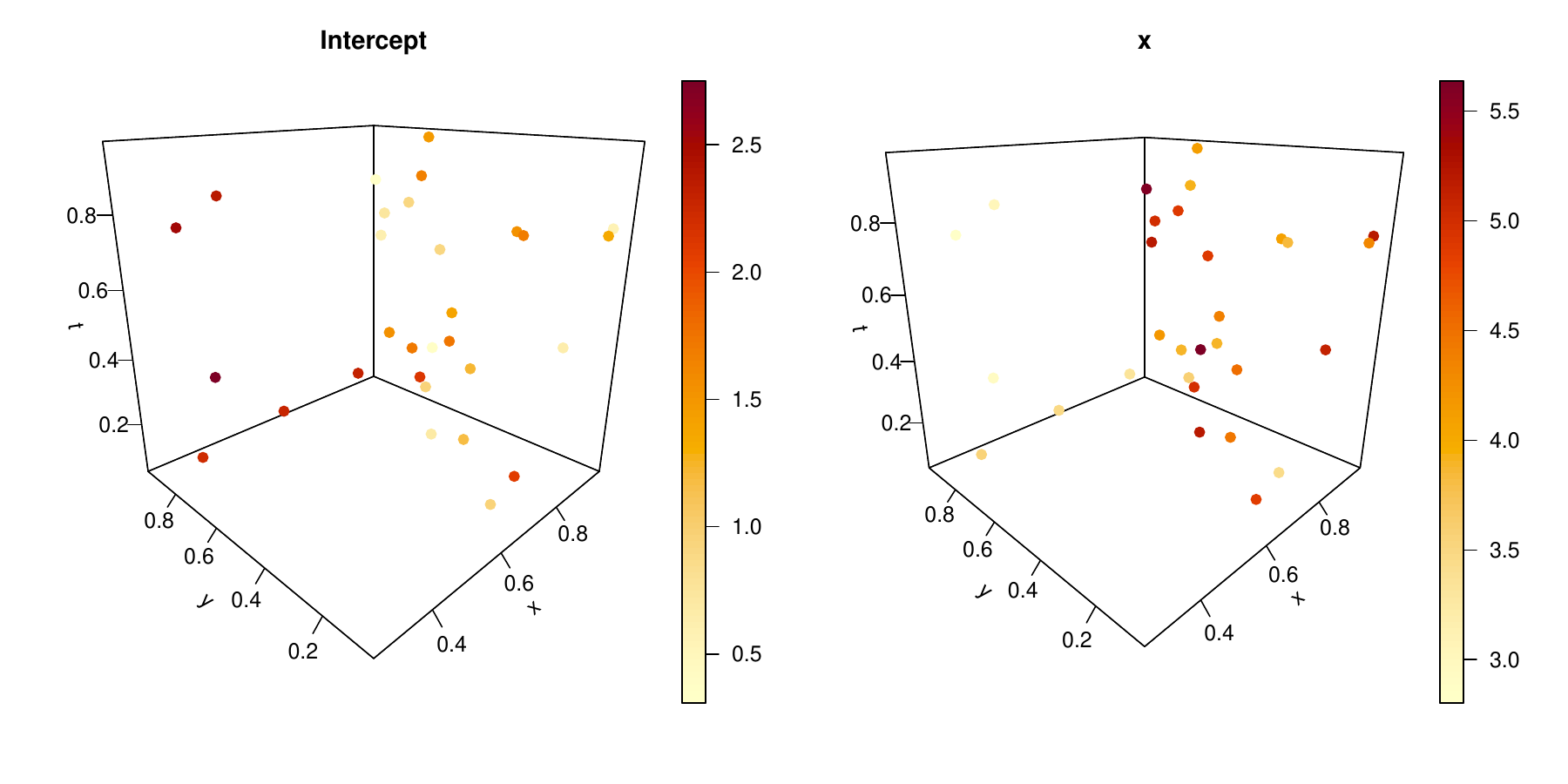}
	\caption{Output of the \proglang{localplot} function applied on a \proglang{locstppm} object.}
	\label{fig:localplot}
\end{figure}

Inference is performed through the fitting of a \proglang{glm} using a localised version of the cubature scheme, firstly introduced
in the spatio-temporal
framework by \cite{d2023locally}. 
Moreover, the \proglang{localplot} function displays the local coefficients overlapped to the observed points in some three-dimensional plots (Figure~\ref{fig:localplot}).

\begin{Sinput}
R> localplot(inh_local)
\end{Sinput}

\begin{figure}[H]
	\centering
	\includegraphics[width=\textwidth]{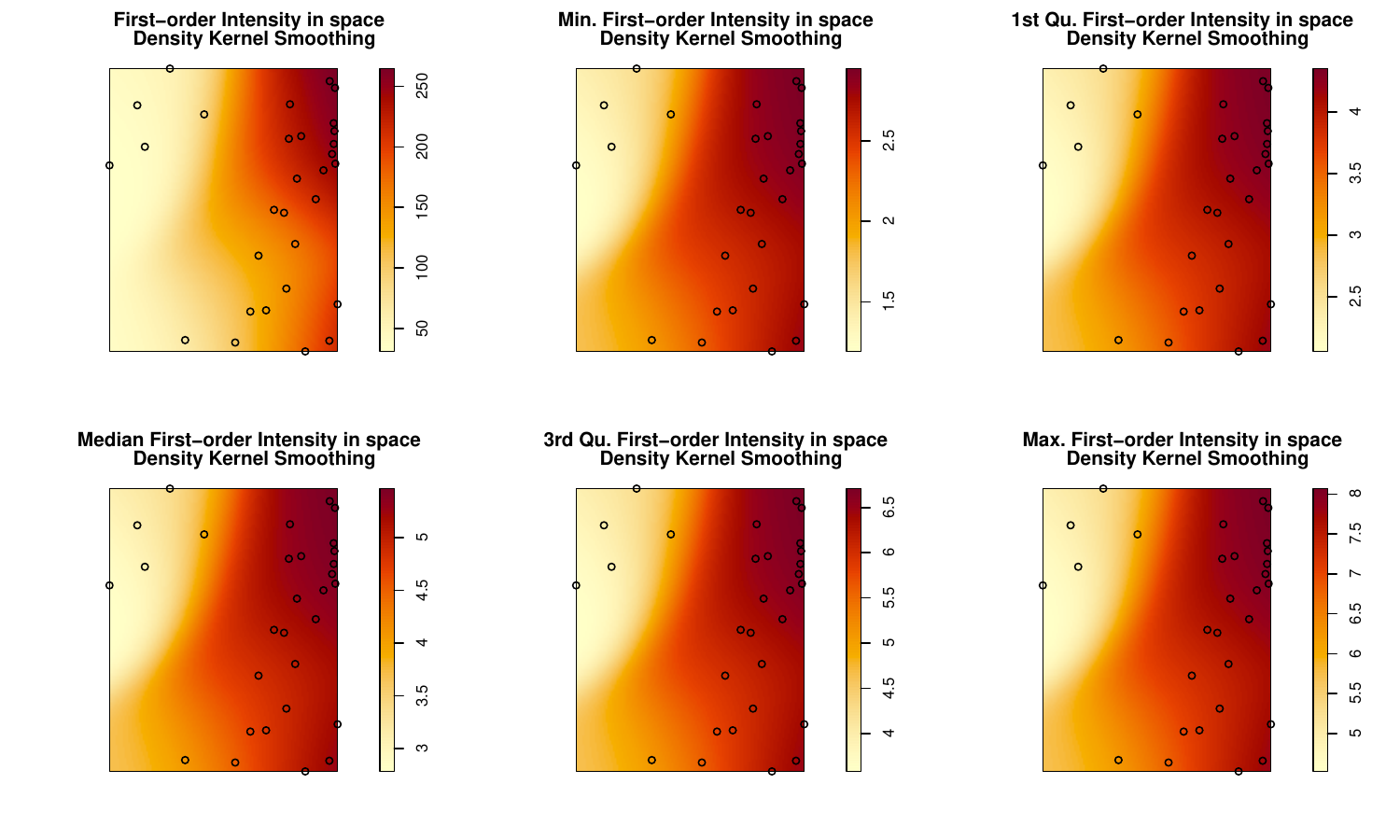}
	\caption{Output of the \proglang{localsummary} function applied on a \proglang{locstppm} object.}
	\label{fig:localsummary}
\end{figure}

Finally, we also implemented the \proglang{localsummary} function, to break up the contribution of the local estimates to the fitted intensity by plotting the overall intensity and the density kernel smoothing of some artificial intensities obtained by imputing the quartiles of the local parameters' distributions (Figure~\ref{fig:localsummary}).

\begin{Sinput}
R> localsummary(inh_local)
\end{Sinput}

\newpage

\subsubsection{Local spatio-temporal log-Gaussian Cox processes}
If the \proglang{second} argument of the   \proglang{stlgcppm} function is set to \proglang{'local'}, it allows to estimate local second-order parameters of a spatio-temporal LGCP, following the \textit{locally weighted minimum contrast} procedure introduced in \cite{d2023locally}.
In particular, we employ the minimum contrast procedure based on the local spatio-temporal pair correlation function \citep{gabriel:rowlingson:diggle:2013} documented in \proglang{LISTAhat} of \pkg{stpp}.
If also \proglang{first} is set to \proglang{'local'}, also the first-order intensity parameters will be fitted locally, obtaining the same achieved by \proglang{locstppm}. 
In the case of local parameters (either first, second-order, or both), the \proglang{print} and \proglang{summary} functions contain information on their distributions.

\begin{Sinput}
R> lgcp2 <- stlgcppm(catsub, second = "local", seed = 2)
R> lgcp2
\end{Sinput}

\begin{Soutput}
Joint minimum contrast fit 
for a log-Gaussian Cox process with 
global first-order intensity and 
local second-order intensity 
--------------------------------------------------
Homogeneous Poisson process 
with Intensity: 0.00849

Estimated coefficients of the first-order intensity: 
(Intercept) 
-4.769 
--------------------------------------------------
Covariance function: separable 

Summary of estimated coefficients of the second-order intensity 
sigma            alpha             beta       
Min.   : 4.867   Min.   :0.1212   Min.   : 7.174  
1st Qu.: 6.740   1st Qu.:0.1776   1st Qu.: 8.528  
Median :13.178   Median :0.3546   Median :12.861  
Mean   :15.638   Mean   :1.0904   Mean   :14.229  
3rd Qu.:18.946   3rd Qu.:1.3206   3rd Qu.:16.433  
Max.   :40.859   Max.   :6.1096   Max.   :31.786  
--------------------------------------------------
Model fitted in 0.88 minutes 
\end{Soutput}

\begin{figure}[H]
	\centering
	\includegraphics[width=.8\textwidth]{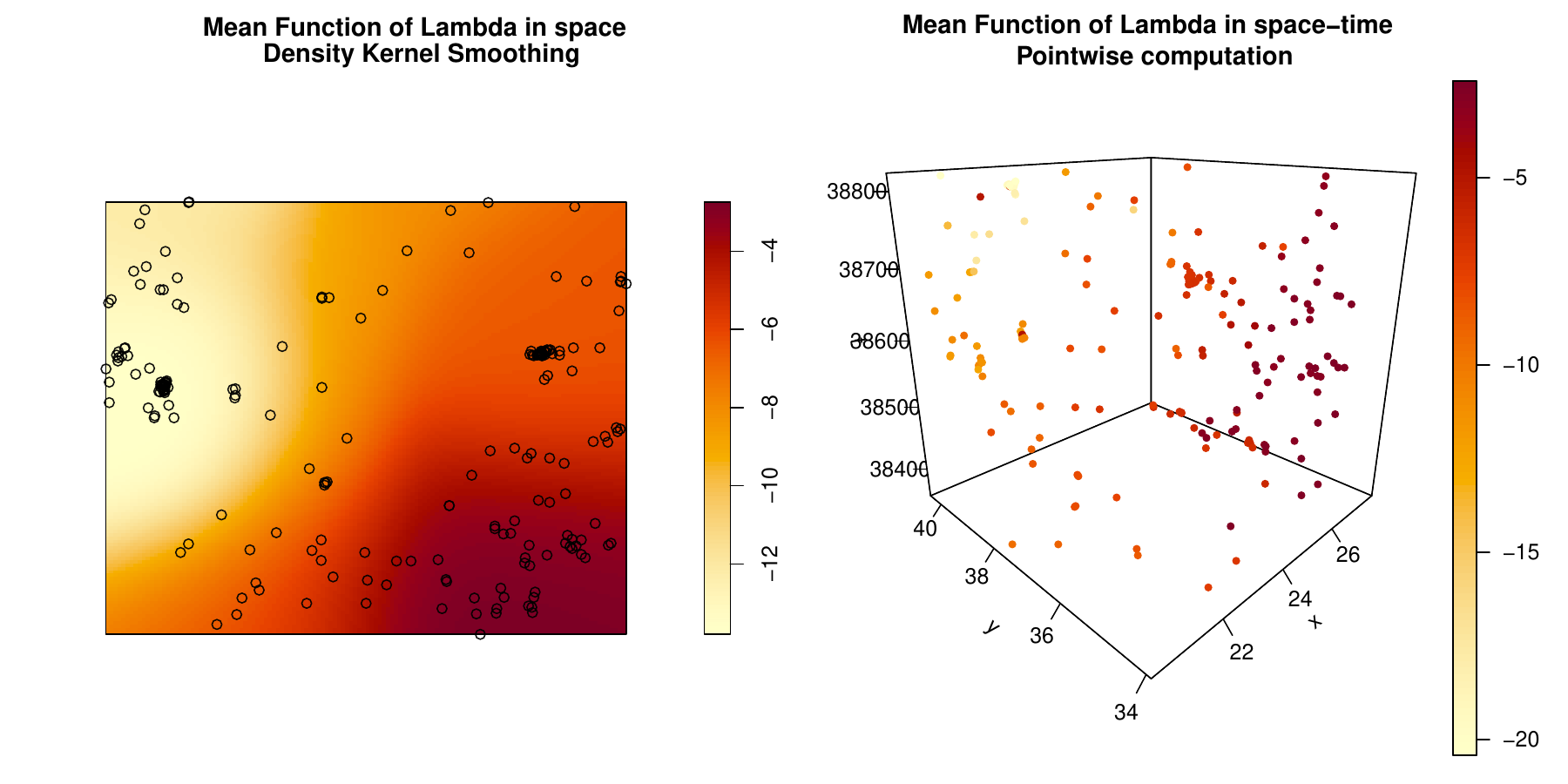}
	\caption{Output of the \proglang{plot} function applied to an estimated LGCP with local covariance parameters.}
	\label{fig:plotl2}
\end{figure}

In the even more specific case of local covariance parameters, the \proglang{plot} function returns the mean of the random intensity, instead of the first-order intensity, displayed both in space (by means of a density kernel smoothing) and in space and time (Figure~\ref{fig:plotl2}).

\begin{Sinput}
R> plot(lgcp2)
\end{Sinput}

\begin{figure}[H]
	\centering
	\includegraphics[width=\textwidth]{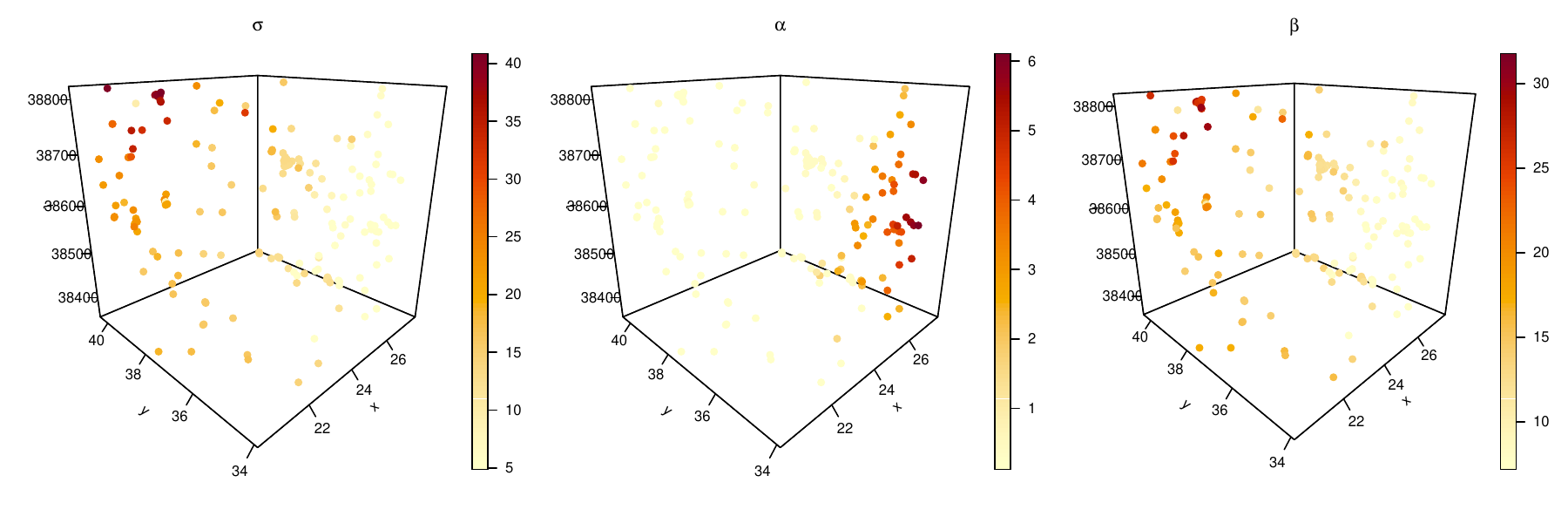}
	\caption{Output of the \proglang{localplot} function applied on a \proglang{stlgcppm} object in the case of local covariance parameters.}
	\label{fig:localsummaryst}
\end{figure}

Finally, the \proglang{localplot} and \proglang{localsummary} functions also work on objects of class \proglang{stlgcppm}, if the LGCP has local first- or second-order fitted parameters. 
In the particular case of local covariance parameters, \proglang{localplot} applied on a \proglang{stlgcppm} object further displays the local estimates of the chosen covariance function (Figure~\ref{fig:localsummaryst}).

\begin{Sinput}
R> localplot(lgcp2)
\end{Sinput}

\section{Diagnostics}\label{sec:diag}
This section is devoted to the presentation of general diagnostic tools based on second-order summary statistics, both globally and locally.

\subsection{Global diagnostics}\label{sec:diag1}
\begin{figure}[H]
	\centering
	\includegraphics[width=\textwidth]{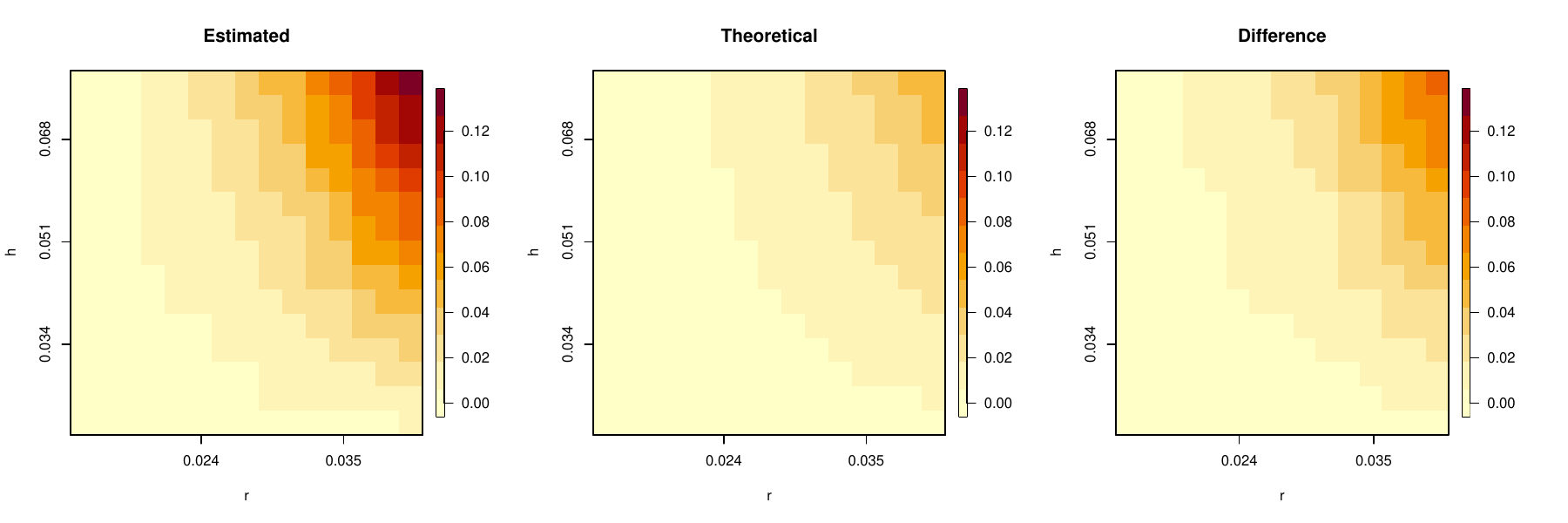}\\
	\includegraphics[width=\textwidth]{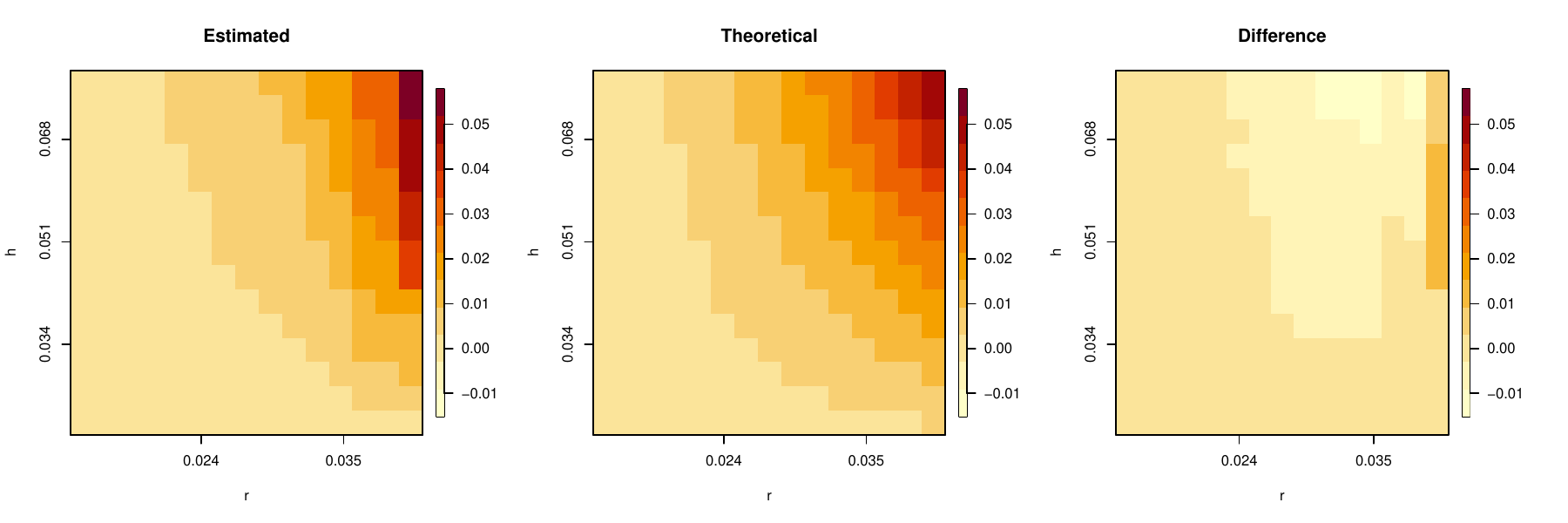}
	\caption{Output of the \proglang{globaldiag} function: $K$-functions weighted by the constant and therefore wrong intensity function (top panels), and $K$-functions weighted by the true intensity function (bottom panels).}
	\label{fig:gdiag}
\end{figure}

The \proglang{globaldiag} function performs global diagnostics of a model fitted for the first-order intensity of a spatio-temporal point pattern, using the spatio-temporal inhomogeneous $K$-function \citep{gabriel2009second} documented by the function \proglang{STIKhat} of the \pkg{stpp} package \citep{stpp}.
It can also perform global diagnostics of a model fitted for the first-order intensity of a spatio-temporal point pattern on a linear network by means of the spatio-temporal inhomogeneous $K$-function on a linear network \citep{moradi2020first} documented by the function \proglang{STLKinhom} of the \pkg{stlnpp} package \citep{stlnpp}.
Both versions return the plots of the inhomogeneous $K$-function weighted by the provided intensity to diagnose, its theoretical value, and their difference (Figure~\ref{fig:gdiag}). 
Next, an example of a simulated point pattern on the unit cube.

\begin{Sinput}
R> set.seed(2)
R> inh <- rstpp(lambda = function(x, y, t, a) {exp(a[1] + a[2] * x)}, 
+    par = c(.3, 6))
R> mod1 <- stppm(inh, formula = ~ 1, seed = 2)
R> mod2 <- stppm(inh, formula = ~ x, seed = 2)
R> (g1 <- globaldiag(mod1))
\end{Sinput}
\begin{Soutput}
Sum of squared differences : 2.036
\end{Soutput}
\begin{Sinput}
R> (g2 <- globaldiag(mod2))
\end{Sinput}
\begin{Soutput}
Sum of squared differences : 0.486
\end{Soutput}
\begin{Sinput}
R> plot(g1)
R> plot(g2)
\end{Sinput}

Figure~\ref{fig:gdiag} displays the result of \proglang{globaldiag} applied to two different fitted intensities: the constant and wrong intensity, and the true one, on the top and bottom panels, respectively. It is evident that the difference between the estimated inhomogeneous $K$-function and its theoretical value is considerably smaller when weighted by the true intensity function.

\subsection{Local diagnostics}\label{sec:diag2}

\begin{figure}[H]
	\centering
	\includegraphics[width=.8\textwidth]{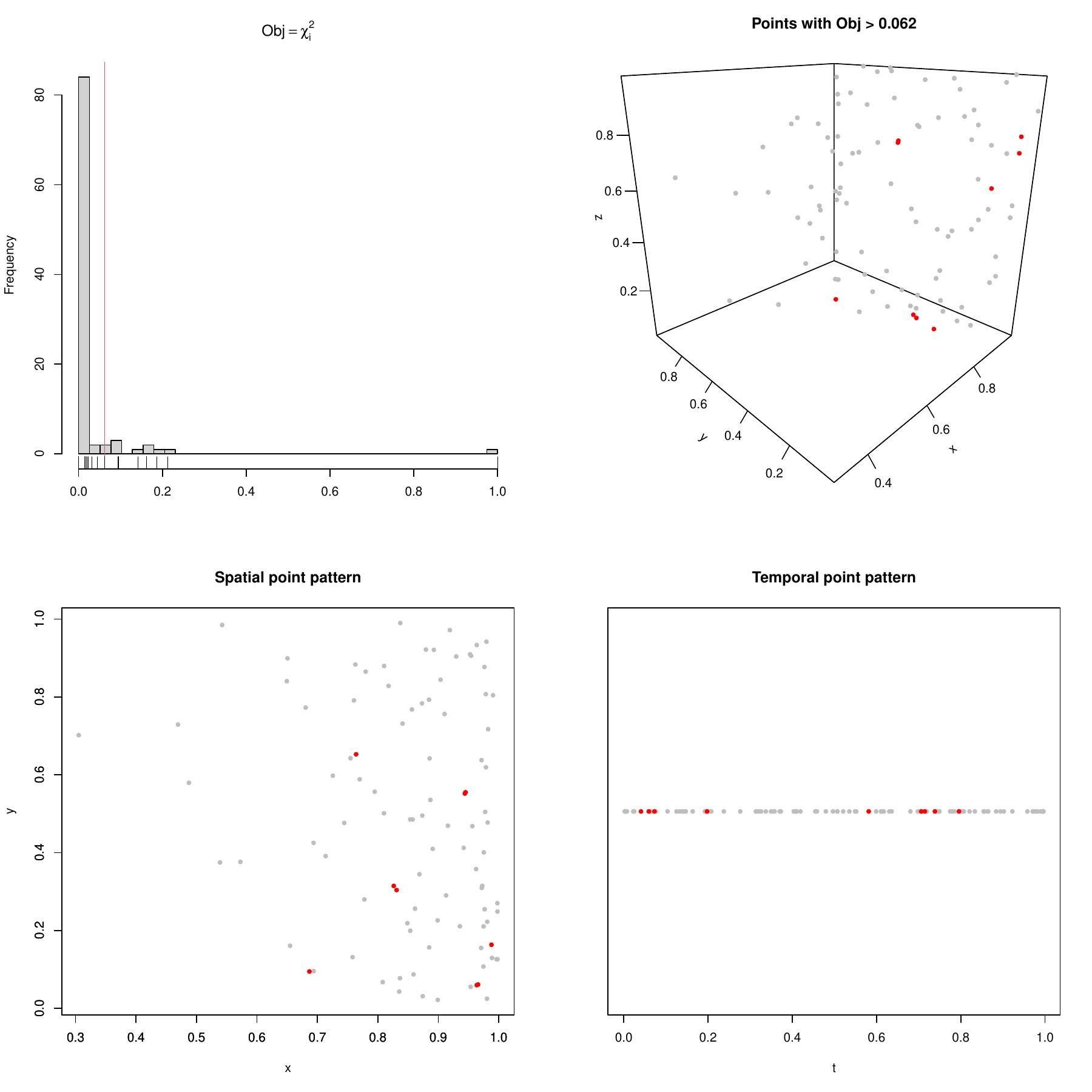}
	\caption{Output of the \proglang{plot.localdiag} function.}
	\label{fig:ldiag}
\end{figure}

\begin{figure}[H]
	\centering
	\includegraphics[width=.8\textwidth]{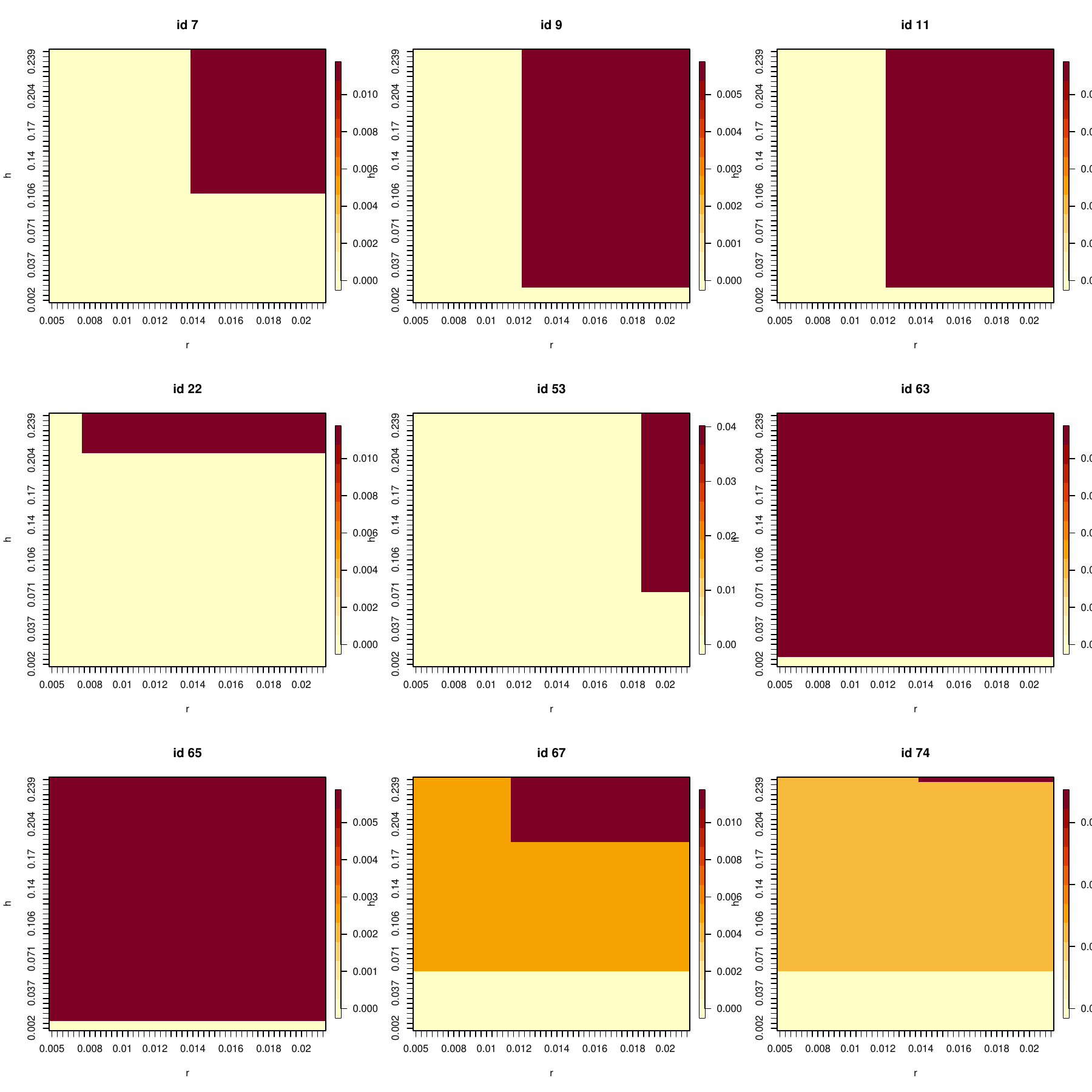}
	\caption{Output of the \proglang{infl} function.}
	\label{fig:ldiag2}
\end{figure}

The \proglang{localdiag} function performs local diagnostics of a model fitted for the first-order intensity of a spatio-temporal point pattern by means of the local spatio-temporal inhomogeneous $K$-functions \citep{adelfio2020some} documented by the function KLISTA of \pkg{stpp}.
It returns the points identified as outlying following the diagnostics procedure on individual points of an observed point pattern, as introduced in \cite{adelfio2020some} and then extended by \cite{dangelo2021local} to the linear network case. \proglang{localdiag} is indeed also able to perform local diagnostics of a model fitted for the first-order intensity of a spatio-temporal point pattern on a linear network by the local spatio-temporal inhomogeneous $K$-functions on linear networks \cite{dangelo2021assessing} documented by the function \proglang{localSTLKinhom} of this package.
The points resulting from the local diagnostic procedure provided by this function can be inspected via the \proglang{plot} (Figure~\ref{fig:ldiag}), \proglang{print}, \proglang{summary}, and \proglang{infl} (Figure~\ref{fig:ldiag2}) functions, as illustrated in the following.

\begin{Sinput}
R> res <- localdiag(inh, mod1$l, p = .9)
R> res
\end{Sinput}

\begin{Soutput}
Points outlying from the 0.9 percentile
of the analysed spatio-temporal point pattern 
--------------------------------------
Analysed pattern X: 97 points 
9 outlying points
\end{Soutput}
\begin{Sinput}
R> plot(res)
\end{Sinput}

In particular, the \proglang{infl} function plots the $K$-functions of all those points identified as outlying by \proglang{localdiag}. Alternatively, one can show only some specific $K$-functions by imputing a vector to the argument \proglang{id}.

\begin{Sinput}
R> infl(res)
\end{Sinput}

\section{Future developments}\label{sec:concl}

The \pkg{stopp} package represents the creation of a toolbox for different spatio-temporal analyses to be performed on observed point patterns, following the growing stream of literature on point process theory. We contribute to the existing literature by framing many of the most widespread methods for the analysis of spatio-temporal point processes into a unique package, which is intended to foster many further extensions. 

The \pkg{stopp} package tools are not exhaustive. 
Some current developments that will be available in future  include the possibility of 
handling irregular spatial windows for purely spatial components and three-dimensional spatial point patterns, fundamental in geology \citep{li2024spherical} and astronomy \citep{babu1996spatial,stoica2007three}. It would be useful to be able to simulate multitype point patterns as well as patterns with intensity depending on external covariates. Also, the next versions of the package could generalise the function \proglang{stppm}, allowing for the inclusion of non-continuous covariates. Alternatives to the inverse-distance weighting for the continuous covariate interpolation could be implemented, including some spatio-temporal smoothing using a Gaussian kernel weighting, which would lead to the Nadaraya-Watson smoother \citep{nadaraya1964estimating,nadaraya1989nonparametric,watson1964smooth}, in addition to the most used kriging \citep{matheron1963principles} and nearest
neighbors interpolation. In addition, the very general Poisson point process model implementation currently available could serve as the basis for the estimation of the first-order intensity function, like the LGCPs. 
Also, other Cox process models relying on the minimum contrast procedure could be implemented, providing the possibility of fitting global and local parameter estimation. 

Moreover, already published research on local characteristics of point processes needing a general software implementation includes: \cite{siino2020spatio,d2023JCGS,d2024advances}.
Regarding alternative fitting methods, it is our intention to give the possibility to use that provided in the recent paper \cite{d2024minimum}.

Currently, the cubature scheme \citep{d2023locally,d2024preprint} is being numerically explored. Indeed, the number of dummy points should be enough for accurate likelihood estimation and experimental results are expected to give guidelines on the number of dummy points to generate. Note that the cubature scheme is already being applied in some work-in-progress analyses of real data in \cite{d2024sds}, for the spatio-temporal analysis of lightning point process data in severe storms, and in \cite{tarantino2024sis}, for the study of a real three-dimensional purely spatial observed point pattern of young stars of the Gaia Archive.

Other interesting implementations which still need some underlying methodological development include the fitting of some general marked point processes with continuous marks, in addition to the already established fitting of general multitype point processes.

\section*{Funding}
This work was supported by the Targeted Research Funds 2024 (FFR 2024) of the University of Palermo (Italy), the Mobilità e Formazione Internazionali - Miur INT project ``Sviluppo di metodologie per processi di punto spazio-temporali marcati funzionali per la previsione probabilistica dei terremoti", and the European Union -  NextGenerationEU, in the framework of the GRINS -Growing Resilient, INclusive and Sustainable project (GRINS PE00000018 – CUP  C93C22005270001).
The views and opinions expressed are solely those of the authors and do not necessarily reflect those of the European Union, nor can the European Union be held responsible for them.


\bibliography{jss5351}

\end{document}